\documentclass[10pt,journal,compsoc]{IEEEtran}
\pdfoutput=1\relax                   
\usepackage{graphicx}                
\DeclareGraphicsExtensions{.pdf,.png,.jpg,.jpeg} 
\usepackage{microtype}                 
\PassOptionsToPackage{warn}{textcomp}  
\usepackage{textcomp}                  
\usepackage{mathptmx}                  
\usepackage{times}                     
\usepackage{cite}                      
\usepackage{tabu}                      
\usepackage{booktabs}                  
\usepackage{verbatim}
\usepackage{amsfonts}
\usepackage{amsmath}
\usepackage{bm}
\usepackage[title]{appendix}
\usepackage[mathscr]{euscript}
\usepackage{algorithm}
\usepackage{algorithmic}
\usepackage{enumitem}
\usepackage{color}
\usepackage[colorlinks,linkcolor=black]{hyperref}
\graphicspath{{figures/}{pictures/}{images/}{./}} 

\begin{document}
%

\title{GNN-Surrogate: A Hierarchical and Adaptive Graph Neural Network for Parameter Space Exploration of Unstructured-Mesh Ocean Simulations}
%
%
%
%

\author{Neng~Shi,
        Jiayi~Xu,
        Skylar~W.~Wurster,
        Hanqi~Guo, \IEEEmembership{Member,~IEEE,}
        Jonathan~Woodring,
        Luke~P.~Van~Roekel, 
        and~Han-Wei~Shen, \IEEEmembership{Member,~IEEE}
\IEEEcompsocitemizethanks{\IEEEcompsocthanksitem Neng Shi, Jiayi Xu, Skylar W. Wurster and Han-Wei Shen are with the Department
of Computer Science and Engineering, The Ohio State University, Columbus, OH, 43210, USA.\protect\\
E-mail: \{shi.1337, xu.2205, wurster.18, shen.94\}@osu.edu
\IEEEcompsocthanksitem Hanqi Guo is with the Mathematics and Computer Science Division, Argonne National Laboratory, Lemont, IL 60439, USA.\protect\\
E-mail: hguo@anl.gov
\IEEEcompsocthanksitem Jonathan Woodring is with the Applied Computer Science Group (CCS-7), Los Alamos National Laboratory, Los Alamos, NM 87544.\protect\\
Email: woodring@lanl.gov
\IEEEcompsocthanksitem Luke P. Van Roekel is with the Fluid Dynamics and Solid Mechanics Group (T-3), Los Alamos National Laboratory, Los Alamos, NM 87544.\protect\\
Email: lvanroekel@lanl.gov
}
}

%
%

\markboth{Journal of \LaTeX\ Class Files,~Vol.~14, No.~8, August~2015}%
{Shell \MakeLowercase{\textit{et al.}}: Bare Demo of IEEEtran.cls for Computer Society Journals}
%



\IEEEtitleabstractindextext{%
\begin{abstract}
We propose GNN-Surrogate, a graph neural network-based surrogate model to explore the parameter space of ocean climate simulations. 
Parameter space exploration is important for domain scientists to understand the influence of input parameters (e.g., wind stress) on the simulation output (e.g., temperature). 
The exploration requires scientists to exhaust the complicated parameter space by running a batch of computationally expensive simulations. 
Our approach improves the efficiency of parameter space exploration with a surrogate model that predicts the simulation outputs accurately and efficiently. 
Specifically, GNN-Surrogate predicts the output field with given simulation parameters so scientists can explore the simulation parameter space with visualizations from user-specified visual mappings. 
Moreover, our graph-based techniques are designed for unstructured meshes, making the exploration of simulation outputs on irregular grids efficient. 
For efficient training, we generate hierarchical graphs and use adaptive resolutions.  
We give quantitative and qualitative evaluations on the MPAS-Ocean simulation to demonstrate the effectiveness and efficiency of GNN-Surrogate. 
Source code is publicly available at  \url{https://github.com/trainsn/GNN-Surrogate}.
\end{abstract}

}

\maketitle

\IEEEdisplaynontitleabstractindextext

%
\IEEEpeerreviewmaketitle

\IEEEraisesectionheading{\section{Introduction}\label{sec:introduction}}

%
%
%
%
\IEEEPARstart{I}{n} oceanography, environment, and climate sciences, scientists usually run ensemble simulations~\cite{wang2018visualization} given different input parameters to perform parameter space analysis and exploration. 
One of the analyses and exploration is to find out the potential relationship between the simulation parameters and outputs~\cite{wang2016multi}. 
For example, Model for Prediction Across Scales-Ocean (MPAS-Ocean) is an ocean model on unstructured grids, and oceanographers control input parameters of interest (e.g., bulk wind stress amplification) to analyze output fields (e.g., temperature) change among different simulation results. 
Scientific visualizations help scientists explore, verify, and summarize differences and similarities between ensemble simulations efficiently and intuitively. 
However, the simulation space exploration requires exhausting the complicated parameter space by running a batch of computationally expensive simulations. 

To make the parameter space exploration efficient, scientists utilize and train a surrogate model by sampling parameter settings from the parameter space. 
Existing surrogate model-based parameter space analyses usually are either image-based (e.g., InSituNet~\cite{he2019insitunet}) or focus on regular grids~\cite{hazarika2019nnva, alden2018using}, which leads to two limitations. 
First, image-based surrogate models visualize the generated simulation data with several predefined visual mappings so that scientists are not able to adjust the setting of visual mappings to find features of interest after the models have been trained.  
Second, existing regular grid-based methods do not directly work for unstructured data since they learn the mapping between the simulation parameters and the raw simulation data on regular grids. 

In this work, to solve the two limitations mentioned above, we present a Graph Neural Network (GNN)-based method to predict raw data on unstructured grids with given simulation parameters. 
We model MPAS-Ocean unstructured meshes as a graph and propose GNN-Surrogate that supports graph operations to learn from MPAS-Ocean simulation output. 
A graph can capture the vertex connectivity and distance information, and thus is one of the common choices to represent unstructured meshes~\cite{perraudin2019deepsphere, defferrard2019deepsphere, pfaff2020learning}. 
For efficient training, we generate hierarchical graphs, where coarse graphs help GNN-Surrogate capture the global phenomena quickly. 
To generate hierarchical graphs, we perform a graph coarsening algorithm. 
Furthermore, we cut the graph hierarchy to reduce the I/O and training computation cost.
Specifically, with the graph hierarchy cutting, GNN-Surrogate adaptively decides which resolutions to use at different locations, depending on how complicated phenomena happen at locations.
Also, the simulation outputs in the training dataset are represented by adaptive resolutions, which supervise the GNN-Surrogate training. 
Given the hierarchical graphs, GNN-Surrogate is an upsampling-convolution generator.
The graph convolution can refine the feature map represented on graphs.
Moreover, its local connectivity and weight sharing scheme allow GNN-Surrogate to avoid over-fitting. 
Domain scientists evaluate our proposed method on MPAS-Ocean~\cite{ringler2013multi}. 

Overall, our workflow is composed of four parts. 
The first part is \textbf{graph hierarchy generation}. Given MPAS-Ocean unstructured meshes, we perform a graph coarsening algorithm to build a graph hierarchy consisting of graphs at different resolution levels.
Then, we cut the graph hierarchy and transform the graphs. 
The second part is \textbf{training data generation}. 
Note that all the simulation outputs in the training dataset are represented by adaptive resolutions, depending on where complex phenomena appear, which can reduce the I/O and training computation cost. 
The third part is \textbf{offline training}. 
We train GNN-Surrogate based on simulation parameters as the input of our model and adaptive resolution output data as the output. 
The fourth part is \textbf{post-hoc exploration and analysis}. 
With a trained GNN-Surrogate, simulation outputs can be predicted with given input parameters and then visualized with existing algorithms. 

GNN-Surrogate can be used for visual analysis on unstructured data, allowing scientists to have a quick preview of simulation outputs given a set of input simulation parameters. 
From the prediction of raw data, various visualization technologies can be applied to let scientists analyze features of interest from different aspects. 
The quick preview makes the visual exploration of the input parameter space more convenient.
Therefore, scientists can efficiently run simulations to analyze the features of interest without going through the entire parameter space. 
Moreover, GNN-Surrogate allows parameter sensitivity analysis for scientists to understand parameter selections better. 

In summary, the main contributions of this paper are twofold:
\setlist{nolistsep}
\begin{itemize}[noitemsep]
  \item We propose GNN-Surrogate to predict simulation outputs given input parameters so that scientists can explore the simulation parameter space with visualizations using user-specified visual mappings.
  
  \item Our graph-based techniques are designed specifically for unstructured meshes, facilitating the efficient exploration of simulation outputs on irregular grids. 
\end{itemize}

\section{Related Work}
\label{section:relate}

In this section, we review related work in parameter space exploration and graph neural networks on unstructured data. 
GNN-Surrogate is also categorized as deep learning for scientific visualization work. 
Readers can check the supplementary material or one survey paper~\cite{wang2021survey} for more details. 

\subsection{Parameter Space Exploration}

We divide existing parameter space exploration work into two categories: (1) traditional methods without surrogate models and (2) surrogate-model based methods. 

Traditional parameter space exploration methods first collect the simulation input and output pairs from ensemble runs, and perform parameter space exploration on the collected pairs. 
In the visualization field, to explore the parameter space of high-dimensional ensemble data, researchers rely on visualization methods such as glyphs~\cite{bock2015visual}, matrices~\cite{poco2014visual}, line charts~\cite{biswas2016visualization}, parallel plots~\cite{obermaier2015visual, wang2016multi}, scatter plots~\cite{orban2018drag, splechtna2015interactive, matkovic2009interactive}, and radial plots~\cite{bruckner2010result, chen2015uncertainty, coffey2013design}.  
The major limitation of these methods is the inability to analyze input parameters that have not been simulated.

Surrogate models, including our GNN-Surrogate, can predict simulation outputs of unseen input parameters for parameter space exploration. 
The surrogate model can be at (1) the image level or (2) the data level.
First, an image-based surrogate model called InSituNet~\cite{he2019insitunet} supports parameter space exploration for ensemble simulations that are visualized in situ. 
Its major limitation is that the simulation output is only visualized with several predefined visual mappings, meaning scientists cannot adjust visual mappings to find features of interest after models have been trained.
Second, researchers have also used different techniques such as machine learning~\cite{hazarika2019nnva, alden2018using} and Gaussian process~\cite{urban2010comparison, erdal2020sampling} to predict raw data using surrogate models. 
Hazarika et al.~\cite{hazarika2019nnva} trained a surrogate model to approximate the yeast cell polarization simulation model in the NNVA system.
Alden et al.~\cite{alden2018using} used a machine learning-based surrogate model to increase users' biological understanding of a simulator of lymphoid tissue organogenesis.  
Urban et al.~\cite{urban2010comparison} proposed a Latin hypercube to improve the quality of a Gaussian process emulator on a simple Earth system model. 
For ensemble-based sensitivity analysis, Erdal et al.~\cite{erdal2020sampling} used a Gaussian process emulator and active subspaces to sample behavioral model parameters.  
However, these data-level methods do not directly work for our unstructured-mesh simulations since they learn the mapping between the simulation parameters and the raw simulation data on regular grids.
Our work is a Graph Neural Network (GNN)-based method specifically designed for unstructured meshes, which makes the exploration of simulation outputs on irregular grids efficient. 

\subsection{Graph Neural Networks on Unstructured Data}
Graph neural networks have been used to learn representations from unstructured data. 
These methods can be divided into (1) spectral methods and (2) spatial methods.

Spectral methods first calculate the eigendecomposition of a graph Laplacian to transform graph signals from the spatial domain to the spectral domain. 
Then the graph convolution is applied to the graph's spectral representation. 
Researchers have applied these methods to spherical data. 
DeepSphere~\cite{perraudin2019deepsphere, defferrard2019deepsphere} is a spherical CNN constructed by representing the sphere as a graph, and spectral CNN operations such as convolution and pooling are defined on it. 
Nevertheless, it is hard for us to exploit edge attribute information of the graph by spectral methods since filters are learned from the spectral domain. 
Thus, we solve our problem by spatial methods.

Spatial methods, on the other hand, define convolutions directly on the graph by grouping neighbors. 
To make use of the edge attributes, researchers designed dynamic edge-conditioned filters. Simonovsky and Komodakis~\cite{simonovsky2017dynamic} used a multi-layer perceptron to compute the convolution kernels given edge labels. 
Valsesia et al.~\cite{valsesia2020learning} represented the edge information by the difference between features on two neighboring nodes.
Lan et al.~\cite{lan2019modeling} proposed Geo-Conv by exploiting the edges' Euclidean geometric information. 
Our kernel is similar to Geo-Conv, while the difference is that we model edges in the spherical polar coordinate rather than Cartesian coordinate since it is more suitable for the MPAS-Ocean datasets. 

\section{MPAS-Ocean Model Background}
\label{section:mpas}

\begin{figure}
  \centering
  \includegraphics[width=1\linewidth]{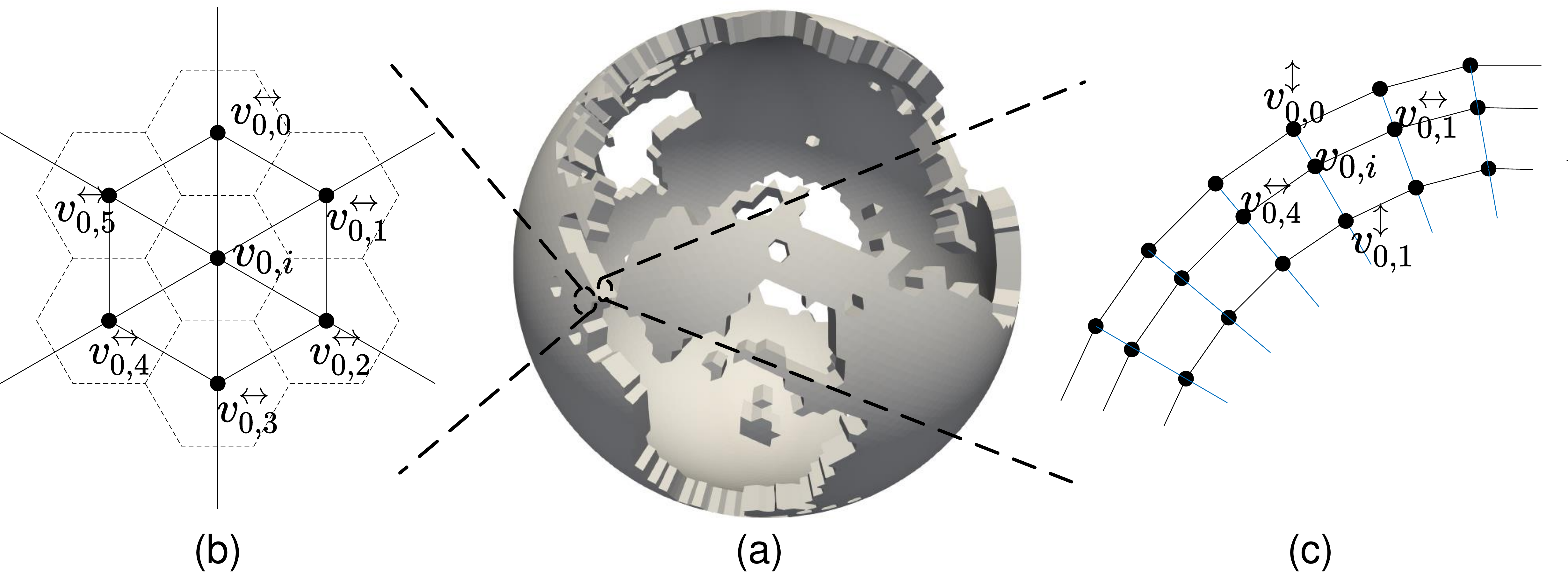}
  \caption{(a) MPAS-Ocean's structure. 
  (b) Horizontal Voronoi polygons. 
  Dashed lines form Voronoi polygons. $v_{0,i}$, $v_{0,0}^\leftrightarrow$, $v_{0,1}^\leftrightarrow$,  $v_{0,2}^\leftrightarrow$,  $v_{0,3}^\leftrightarrow$,  $v_{0,4}^\leftrightarrow$,  $v_{0,5}^\leftrightarrow$ are Voronoi polygon cell centers. 
  Solid lines link cell centers. 
  (c) Cross section. 
  $v_{0,0}^\updownarrow$, $v_{0,1}^\updownarrow$ are $v_{0,i}$'s vertical neighbors. 
  Blue lines link vertical neighbors.}
  \label{fig:cell2graph}
\end{figure}

MPAS-Ocean~\cite{ringler2013multi} is a model for describing and evaluating the ocean. 
By exploiting unstructured meshes, MPAS-Ocean is particularly suitable for regionally enhancing the resolution without influencing the global simulation quality.

\begin{figure}
  \centering
  \includegraphics[width=1\linewidth]{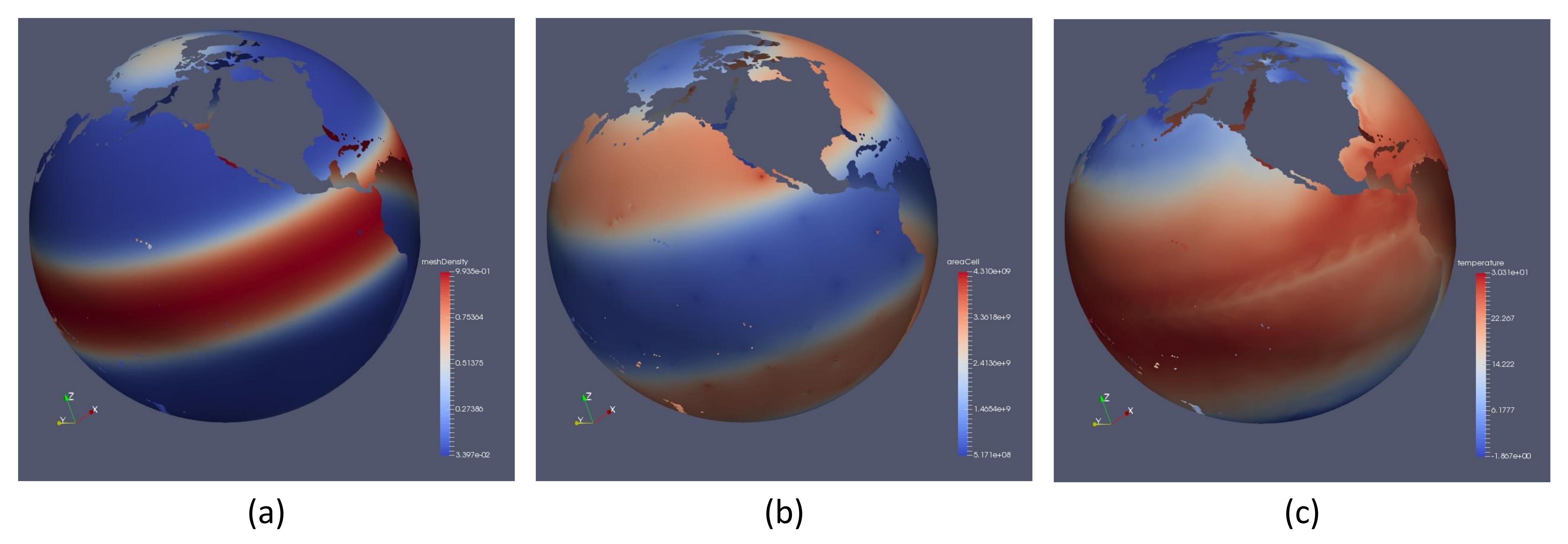}
  \caption{Scalar fields. (a)Mesh Density. (b)Cell Size. (c)Temperature.}
  \label{fig:adaptive}
\end{figure}

MPAS-Ocean's mesh structure is shown in Figure~\ref{fig:cell2graph}. In earth science, the directions consistent with and perpendicular to the gravity are defined as \textit{vertical} and \textit{horizontal}, respectively. 
The oceans are divided both horizontally and vertically by oceanographers. 
Scientists try to solve issues related to horizontal and vertical discretization by MPAS Ocean. 
The \textit{horizontal} grids depend on Spherical Centroidal Voronoi Tessellations (SCVTs), i.e., a spherical surface is composed of Voronoi regions. 
Some variables of interest (e.g., temperature) are stored at the Voronoi polygon cell centers.
Note that in one horizontal spherical layer, the Voronoi region distributions on the sphere are based on a user-defined mesh-density function, as shown in Figure~\ref{fig:adaptive}(a).
For example, for the EC60to30~\cite{petersen2018ec60to30} mesh, the grids' cell sizes vary from 30km to 60km. 
As shown in Figure~\ref{fig:adaptive}(b), scientists want to use high resolution in equatorial and polar regions because some interesting natural phenomenon happens there such as eastern equatorial Pacific cold tongue, which can be reflected from the temperature field in Figure~\ref{fig:adaptive}(c). 

Vertically, to model the 3D ocean, scientists scale and copy the Voronoi polygons to different depth levels and build spherical layers perpendicular to the Earth’s surface. Scientists sample densely near the ocean surface to model complex phenomenons. 

\section{GNN-Surrogate Overview}
\label{section:overview}

\begin{figure*}
  \centering
  \includegraphics[width=\linewidth]{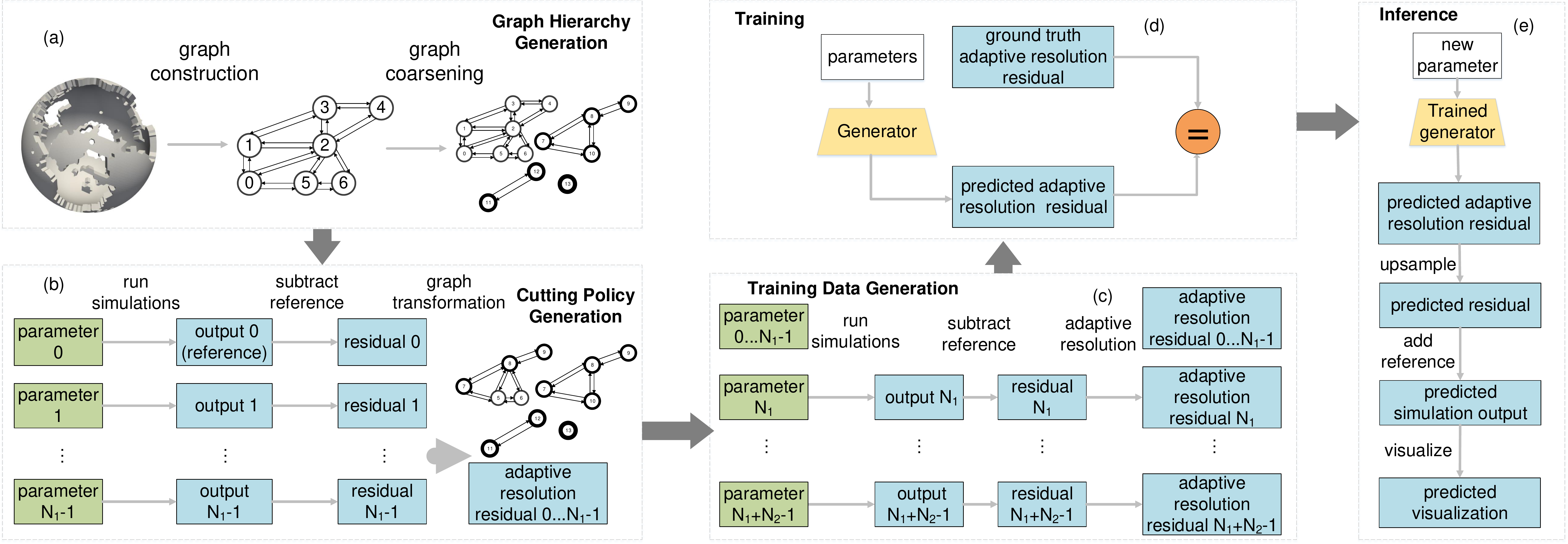}
  \caption{Workflow of our approach. (a) Given the MPAS-Ocean mesh structure, a corresponding graph hierarchy is generated. 
  (b) A few simulations are run for generating the graph hierarchy cutting policy. 
  The cutting policy is used to guide representing the simulation output with adaptive resolutions. 
  (c) Another batch of ensemble simulations is run for collecting the training data. 
  (d) A deep surrogate model (i.e., GNN-Surrogate) is trained based on the generated training dataset. 
  (e) In the inference stage, GNN-Surrogate is used to predict the simulation output.
  The predicted simulation output can be visualized later for parameter space exploration.}
  \label{fig:overview}
\end{figure*}

Our goal is to develop a model to support parameter exploration and visualization of ocean ensemble simulations with some input simulation parameters.  
We achieve this by creating a surrogate model predicting simulation outputs given simulation parameters. 
As mentioned in Section~\ref{section:mpas}, MPAS is successful because it exploits unstructured meshes. 
To support machine learning on unstructured meshes, we design a graph neural network (GNN)-based model, GNN-Surrogate, to directly predict the ocean data presented by unstructured grids. 

Figure~\ref{fig:overview} provides the workflow of our approach. 
GNN-Surrogate is an upsampling-convolution generator, which requires graphs at different resolution levels. 
We construct a level-zero graph ($G_0$) given the full resolution of MPAS-Ocean's unstructured meshes (Section~\ref{subsection:construction}) and perform a graph coarsening algorithm to build a graph hierarchy consisting of graphs at different resolution levels ($G_1,\ G_2,\  \ldots$) (Section~\ref{subsection:coarsen}). 
Due to the GPU memory size constraint, GPU memory may not be sufficient to hold the intermediate feature maps of full-resolution data. 


Our solution to the GPU memory size constraint problem is to use adaptive data resolutions. 
We first run a few simulations with random input simulation parameters to gather the graph hierarchy cutting policy.  
The cutting policy is made by finding where complex phenomena appear in the simulation outputs.
We discuss the number of simulation runs required for the cutting policy in the supplementary material. 
Following the cutting policy, we cut the graph hierarchy and transform the hierarchical graphs, which can reduce the I/O and training computation cost. (Section~\ref{subsection:cut}). 
The transformed graphs are used as templates to generate adaptive resolution outputs for future simulation runs.  (Section~\ref{subsection:collect}). 
Then, GNN-Surrogate that consists of specifically designed graph convolution (Section~\ref{subsection:convolution}) and upsampling (Section~\ref{subsection:upsample}) operators, is trained to learn the mapping from simulation parameters to simulation outputs with adaptive resolutions (Section~\ref{subsection:network}).  
GNN-Surrogate can first create low-resolution feature maps containing the global data information and represent them in coarser graphs with the transformed graph hierarchy. 
Then GNN-Surrogate can refine the feature maps until the feature maps become a full resolution. 
Finally, during post-hoc exploration and analysis, with a trained GNN-Surrogate, scientists can predict a simulation output given a new input parameter setting and then use existing visualization algorithms to generate visualization images and perform further analysis and evaluation. 

\section{Data Structure: Hierarchical Graphs for Data with Adaptive Resolutions} 
\label{section:hiegraphs}

Our goal is to learn a function $F$ that maps a set of input simulation parameters $P_{sim}$ to the corresponding simulation output $S$ used for visualization. 
The function can be defined as: $F(P_{sim}) \rightarrow S$.
Training a deep model for high-resolution MPAS-Ocean data can be prohibitively expensive.
To reduce the computation and memory cost, we decompose the function as $F(P_{sim}) = F_{ar \rightarrow s}(F_{p \rightarrow ar}(P_{sim}))$,
where $F_{p \rightarrow ar}$ is the function maps the input simulation parameter to the output represented with adaptive resolutions, and $F_{ar \rightarrow s}$ converts the adaptive resolution result to the predicted simulation output with full resolution. 

In this section, we build hierarchical graphs ($G_0, \ G_1,\ G_2,\  \ldots$) to represent outputs with adaptive resolutions. 
The graph $G_0$ is built on the meshes with full resolution, which is presented next. 

\subsection{Edge-weighted Graph Construction}
\label{subsection:construction}

We build a directed graph $G_0 = (V_0, E_0)$ of level zero from MPAS-Ocean unstructured geodesic grids, where $V_0$ is the node set, $E_0$ is the edge set. 
A variable of interest (cell-centered) such as temperature is defined on a node $v_{0,i}$. 
Every node $v_{0,i}$ is attached with its Cartesian coordinate $(x_{0,i}, y_{0,i}, z_{0,i})$ and spherical coordinate  $(r_{0,i}, \theta_{0,i}, \phi_{0,i})$. 

$E_0$ can be decomposed into two groups, horizontal edges $E_0^\leftrightarrow$ and vertical edges $E_0^\updownarrow$, based on geographic information of edges, where $E_0 = E_0^\leftrightarrow \cup  E_0^\updownarrow$. 
Note that \textit{horizontal} and \textit{vertical} are the directions  perpendicular to and consistent with the  gravity, respectively, as illustrated in Figure~\ref{fig:cell2graph}. 
We define horizontal and vertical edges separately because the horizontal and vertical resolutions are not on the same scale. 
A node $v_{0,i}$ is connected with its horizontal neighbors across the cell walls, which are denoted as graph nodes $v_{0,j}^\leftrightarrow$, and $(v_{0,i}, v_{0,j}^\leftrightarrow) \in E_0^\leftrightarrow$ is formed as horizontal edges, as shown in Figure~\ref{fig:cell2graph}(b).
Most nodes are connected with six horizontal neighbors, but many nodes near the continents have $1 \sim 5$ horizontal neighbors. 
Also, $v_{0,i}$ is connected to no more than two vertical neighbors $v_{0,j}^\updownarrow$, and $(v_{0,i}, v_{0,j}^\updownarrow) \in E_0^\updownarrow$ is formed as vertical edges, as shown in Figure~\ref{fig:cell2graph}(c).

To describe the geographical relationships, we first compute the distance between two edge-connected nodes, where a shorter distance means the two nodes strongly influence each other. 
The distance is measured as follows, 
\begin{equation}\nonumber
d(v_{0, i}, v_{0, j})=
\left\{
    \begin{array}{lr}
    d^\leftrightarrow(v_{0, i}, v_{0, j}) & \textrm{if} \ (v_{0, i}, v_{0, j}) \in E_0^\leftrightarrow, \\
    d^\updownarrow(v_{0, i}, v_{0, j}) & \textrm{if} \ (v_{0, i}, v_{0, j}) \in E_0^\updownarrow, \\
    +\infty & \textrm{otherwise}, 
    \end{array}
\right.
\end{equation}
where $d^\leftrightarrow(v_{0, i}, v_{0, j})$ is the great-circle distance~\cite{feld1940plane} and $d^\updownarrow(v_{0, i}, v_{0, j})$ is the Euclidean distance. 

Inspired by DeepSphere\cite{perraudin2019deepsphere}, we formulate the geographical relationships between two edge-connected nodes  by weight of the edge, which is defined as 
\begin{equation}
w_0(v_{0, i}, v_{0, j}) = 
\left\{
             \begin{array}{lr}
             exp\left(-\dfrac{(d^\leftrightarrow(v_{0, i}, v_{0, j}))^2}{4\rho_0^\leftrightarrow}\right) & \textrm{if} \ (v_{0, i}, v_{0, j}) \in E_0^\leftrightarrow, \\
             exp\left(-\dfrac{(d^\updownarrow(v_{0, i}, v_{0, j}))^2}{4\rho_0^\updownarrow}\right) & \textrm{if} \ (v_{0, i}, v_{0, j}) \in E_0^\updownarrow, \\
             0 & \textrm{otherwise}, 
             \end{array}
\right.
\end{equation}
A higher weight comes from a shorter distance, indicating that two nodes have a stronger relationship, and 
\begin{eqnarray}\nonumber
\rho_0^\leftrightarrow = \dfrac{1}{\left|E_0^\leftrightarrow\right|} \sum_{(v_{0, i}, v_{0, j}) \in E_0^\leftrightarrow} (d^\leftrightarrow(v_{0, i}, v_{0, j}))^2 , \\
\nonumber \rho_0^\updownarrow = \dfrac{1}{\left|E_0^\updownarrow\right|} \sum_{(v_{0, i}, v_{0, j}) \in E_0^\updownarrow} (d^\updownarrow(v_{0, i}, v_{0, j}))^2
\end{eqnarray}
are the average distance square of all the horizontal and vertical edges, respectively. 
This weighting scheme performs well empirically, demonstrated in both DeepSphere~\cite{perraudin2019deepsphere} and our experiments. 

\begin{figure}
  \centering
  \includegraphics[width=1\linewidth]{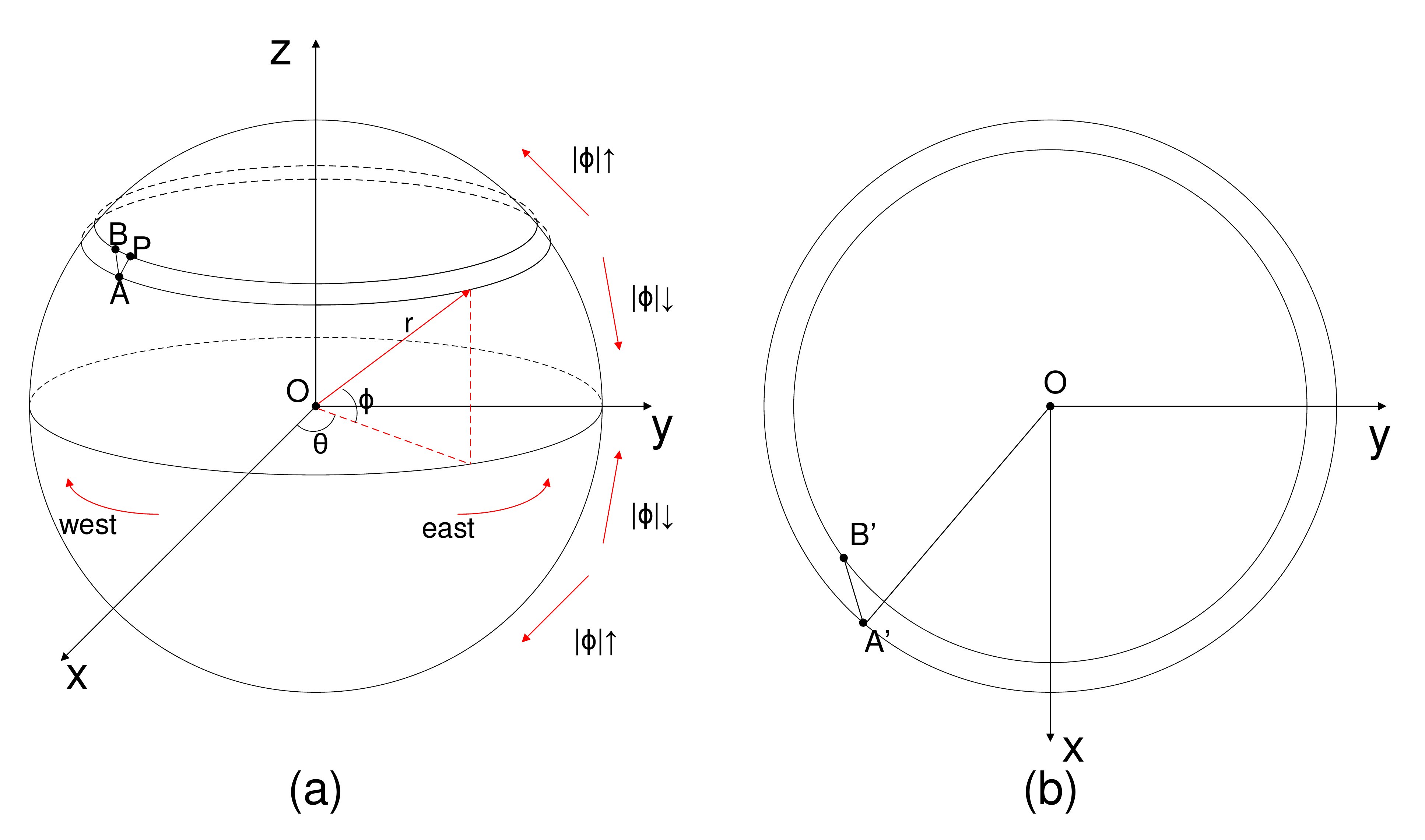}
  \caption{Horizontal Edge Attribute. (a) The horizontal edge $(A, B)$ and its $|\phi|\uparrow$ component $\overset{\frown} {AP}$. (b) The platform. $A^\prime B^\prime$ is edge $(A, B)$'s west component. }
  \label{fig:decompose}
\end{figure}

To further describe the geographical directional relationship between two edge-connected nodes, we define six directions: 
toward high latitude ($|\phi|\uparrow$), toward low latitude ($|\phi|\downarrow$), westward (West), eastward (East), Shallower ($r\uparrow$), and Deeper ($r\downarrow$), which is illustrated in Figure~\ref{fig:decompose}. 
The six directions are needed because the geographical directional relationship matters for planetary signals such as ocean data. 
East and West are earth rotation and counter-earth rotation direction. $|\phi|\uparrow$ and $r\downarrow$ are the directions where solar radiation energy decreases horizontally and vertically, respectively, and the opposite for direction $|\phi|\downarrow$ and $r\uparrow$. 
A edge direction set $\mathscr{D} = \{|\phi|\uparrow, |\phi|\downarrow, West, East, r\uparrow, r\downarrow\}$ is built to describe every edge between two connected nodes. 
Then, the attribute of an edge $(v_{0, i}, v_{0, j}) \in E_0$ is represented as an edge attribute vector $\bm {\Gamma_0}(v_{0, i}, v_{0, j})$. 
To calculate $\bm{\Gamma_0}(v_{0, i}, v_{0, j})$, we first define offset between two nodes $v_{0, i}$ and $v_{0, j}$ as 
\begin{eqnarray}
  \nonumber \delta_0^x = x_{0,j} - x_{0,i}, \
  \delta_0^y = y_{0,j} - y_{0,i}, \ 
 \delta_0^r = r_{0,j} - r_{0,i}, \\
 \nonumber \delta_0^{lat} = |\phi_{0,j}| - |\phi_{0,i}|, \ 
 \delta_0^{lng} = \dfrac{||(\delta_0^x, \delta_0^y)||_2}{r_{0,i}}, \ 
 \delta_0^h = |\delta_0^{lat}| + |\delta_0^{lng}|,
\end{eqnarray}
where $||(\delta_0^x, \delta_0^y)||_2$ is the 2-norm of $(\delta_0^x, \delta_0^y)$.   
Then, $\bm{\Gamma_0}(v_{0, i}, v_{0, j}) = (\gamma_0^{|\phi|\uparrow}, \gamma_0^{|\phi|\downarrow}, \gamma_0^{west}, \gamma_0^{east}, \gamma_0^{r\uparrow}, \gamma_0^{r\downarrow})^\mathsf{T}$ is initialized by all zeros and updated by: 
\begin{equation}
\begin{array}{lr}
\gamma_0^{|\phi|\uparrow} = w_0(v_{0, i}, v_{0, j}) \cdot \dfrac{|\delta_0^{lat}|}{\delta_0^h} & \textrm{if} \  \delta_0^{lat} > 0, \\
\gamma_0^{|\phi|\downarrow} = w_0(v_{0, i}, v_{0, j}) \cdot \dfrac{|\delta_0^{lat}|}{\delta_0^h} & \textrm{if} \ \delta_0^{lat} < 0, \\
\gamma_0^{west} = w_0(v_{0, i}, v_{0, j}) \cdot \dfrac{|\delta_0^{lng}|}{\delta_0^h} & \textrm{if} \ (x_{0,i}, y_{0,i})^\mathsf{T} \times (\delta_0^x, \delta_0^y)^\mathsf{T} > 0, \\
\gamma_0^{east} = w_0(v_{0, i}, v_{0, j}) \cdot \dfrac{|\delta_0^{lng}|}{\delta_0^h} & \textrm{if} \ (x_{0,i}, y_{0,i})^\mathsf{T} \times (\delta_0^x, \delta_0^y)^\mathsf{T} < 0, \\
\gamma_0^{r\uparrow} = w_0(v_{0, i}, v_{0, j}) & \textrm{if} \ \delta_0^r > 0, \\
\gamma_0^{r\downarrow} = w_0(v_{0, i}, v_{0, j}) & \textrm{if} \ \delta_0^r < 0,
\end{array}
\end{equation}
where $\times$ is the cross product, and we guarantee $\gamma_0^{|\phi|\uparrow} + \gamma_0^{|\phi|\downarrow} + \gamma_0^{west} + \gamma_0^{east} + \gamma_0^{r\uparrow} + \gamma_0^{r\downarrow} = w_0(v_{0, i}, v_{0, j})$ to ensure the sum of vector components is the edge weight. 

For example, in Figure~\ref{fig:decompose}, for one horizontal edge $(v_A,v_B)$, the edge attribute vector is 
\begin{align*}
& \bm{\Gamma_0}(v_A,v_B) \\ 
& = 
(w_0(v_A, v_B) \cdot \dfrac{|\overset{\frown} {AP}|}{|\overset{\frown} {AP}|+ |A^\prime B^\prime|}, 0, w_0(v_A, v_B) \cdot \dfrac{|A^\prime B^\prime|}{|\overset{\frown} {AP}|+ |A^\prime B^\prime|}, 0, 0, 0)^\mathsf{T}.
\end{align*}

\subsection{Graph Hierarchy Generation for Efficient Training}
\label{subsection:coarsen}

We generate a graph hierarchy for GNN-Surrogate efficient training by using a graph coarsening algorithm. 
The graph coarsening operation receives an input graph $G_l$ and outputs the next level coarser graph $G_{l+1}$ which has fewer nodes and edges but preserves the input graph's topological structures.
We repeat the coarsening process until we are left with a single node in the coarsened graph. 
In this work, we use two simple and efficient edge matching algorithms~\cite{dhillon2007weighted} on horizontal and vertical edges, respectively.
We want to avoid mixing horizontal and vertical edges because, in coarser graphs, we want to use the same structure for every spherical layer and have the edges in the coarser graphs be either horizontal or vertical, but not mixed, as in the original graph $G_0$. 

The horizontal matching works as follows: given a graph $G_l$, all the nodes are initially marked as unvisited. 
Starting from the top horizontal layer, we match a node $v_{l,i}$ with its cloest horizontal neighbor $v_{l,j}$(the one that maximizes $w_l(v_{l, i}, v_{l, j})$) and generate one super-node $v_{l+1, k}$ in the next coarser graph $G_{l+1}$. 
The super-node's position is the two matched nodes' great circle midpoint. 
If there is no unvisited horizontal neighbor, we copy $v_{l, i}$ itself to the the coarser graph as the super-node $v_{l+1, k}$ (such as $G_0$ node 4 in Figure~\ref{fig:coarsen}(a)). 
The one singleton or two matched nodes are then marked as visited. We repeat the matching process until all nodes in the top horizontal layer are marked as visited.
Figure~\ref{fig:coarsen}(a) illustrates the process. 
We then apply the same strategy for nodes in other horizontal layers. 
Given two super-nodes $v_{l+1,i}$ and $v_{l+1,j}$ in $G_{l+1}$, if any of $v_{l+1,i}$'s children and any of $v_{l+1,j}$'s children are neighbors, $v_{l+1,i}$ and $v_{l+1,j}$ are considered as neighbors and an edge is linked between them. 
We use the same edge attribute generation and weighting scheme for $G_{l+1}$ as for the initial graph $G_0$ described in Section~\ref{subsection:construction}. 

For vertical matching, we match nodes in the odd horizontal layers with their neighbors in the deeper layer. 
If the number of horizontal layers is odd, the nodes in the deepest horizontal layer would be singletons. 
The edge forming, weighting scheme, and edge attribute generation are the same as the horizontal matching. 
Both edge matching algorithms' time complexity are $O(\left|V_l\right|)$.  

During coarsening, starting from the original graph $G_0$, we alternately perform horizontal and vertical edge matching to generate graphs at different resolution levels $G_1, G_2, \ldots, G_{L-1}$. 
The graph coarsening details can be found in the supplementary material. 
We store all the graphs into a list, which is called the graph hierarchy. 
We also store the parent-children links in the graph hierarchy and construct a graph hierarchical tree (GHT). 
Figure~\ref{fig:coarsen}(b) shows a GHT example.     
The graph hierarchy is used for a final transformation process to decide the graphs used in our GNN-Surrogate, described in the next section. 

\subsection{Hierarchical Tree Cutting}
\label{subsection:cut}

\begin{figure*}
  \centering
  \includegraphics[width=\linewidth]{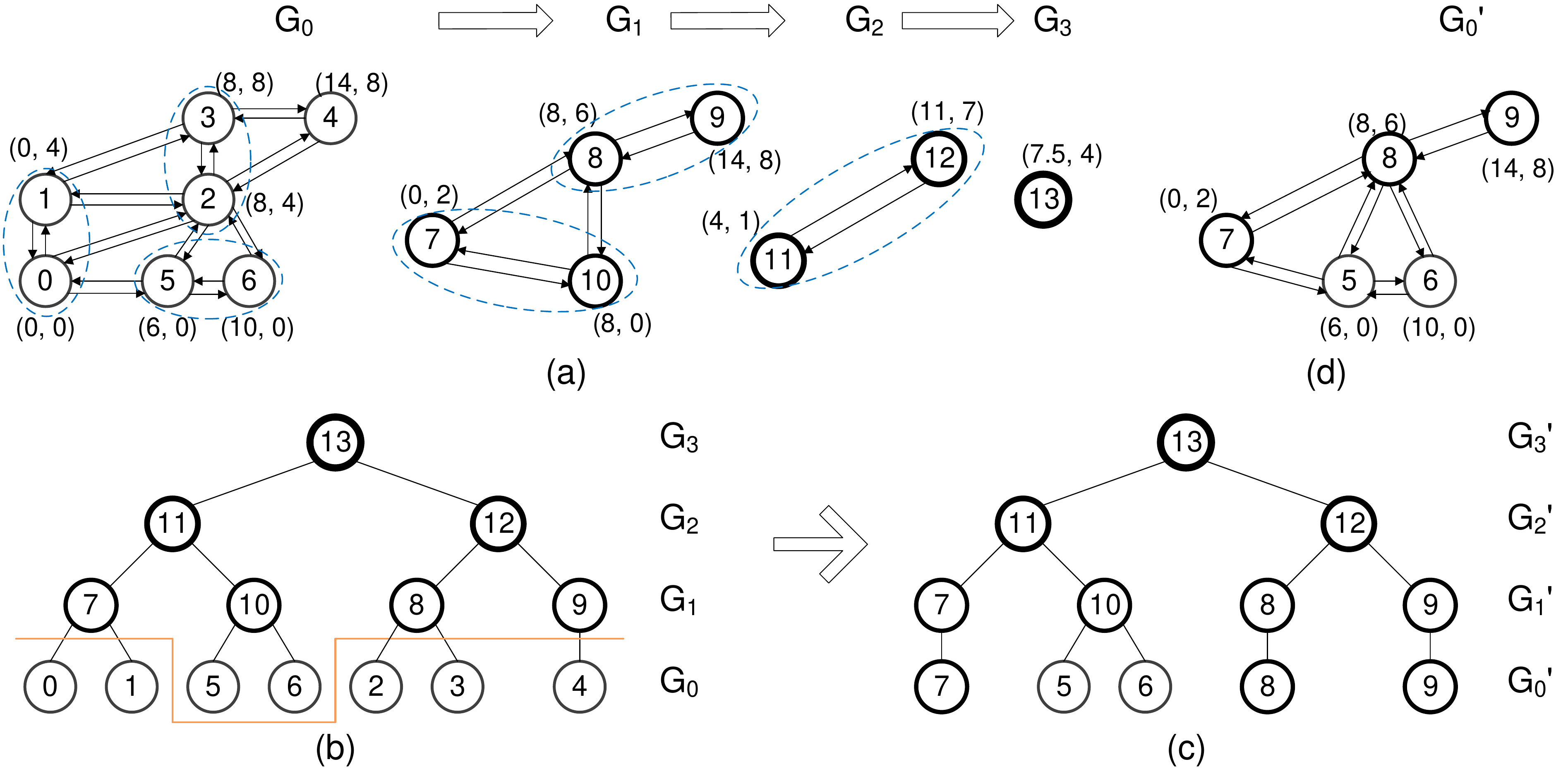}
  \caption{Pipeline of the graph hierarchy generation and graph transformation. 
  (a) Example of the graph coarsening by horizontal edge matching. 
  Given the original graph $G_0$, horizontal matching is performed for a coarser graph until $G_3$, a graph with an isolated point, is reached. 
  $G_0$, $G_1$, $G_2$, $G_3$ form a graph hierarchy. 
  (b) The graph hierarchical tree (GHT) is created by storing the parent-children links in the graph hierarchy. 
  A GHT cut is generated. In the example, nodes below the orange curve are cut. 
  (c) Tree view of the transformed graph hierarchy. 
  The transformation is performed based on the cutting. 
  $G_0$ is affected in the transformation. 
  (d) $G_0^\prime$ is transformed from $G_0$. }
  \label{fig:coarsen}
\end{figure*}

\begin{figure}
  \centering
  \includegraphics[width=\linewidth]{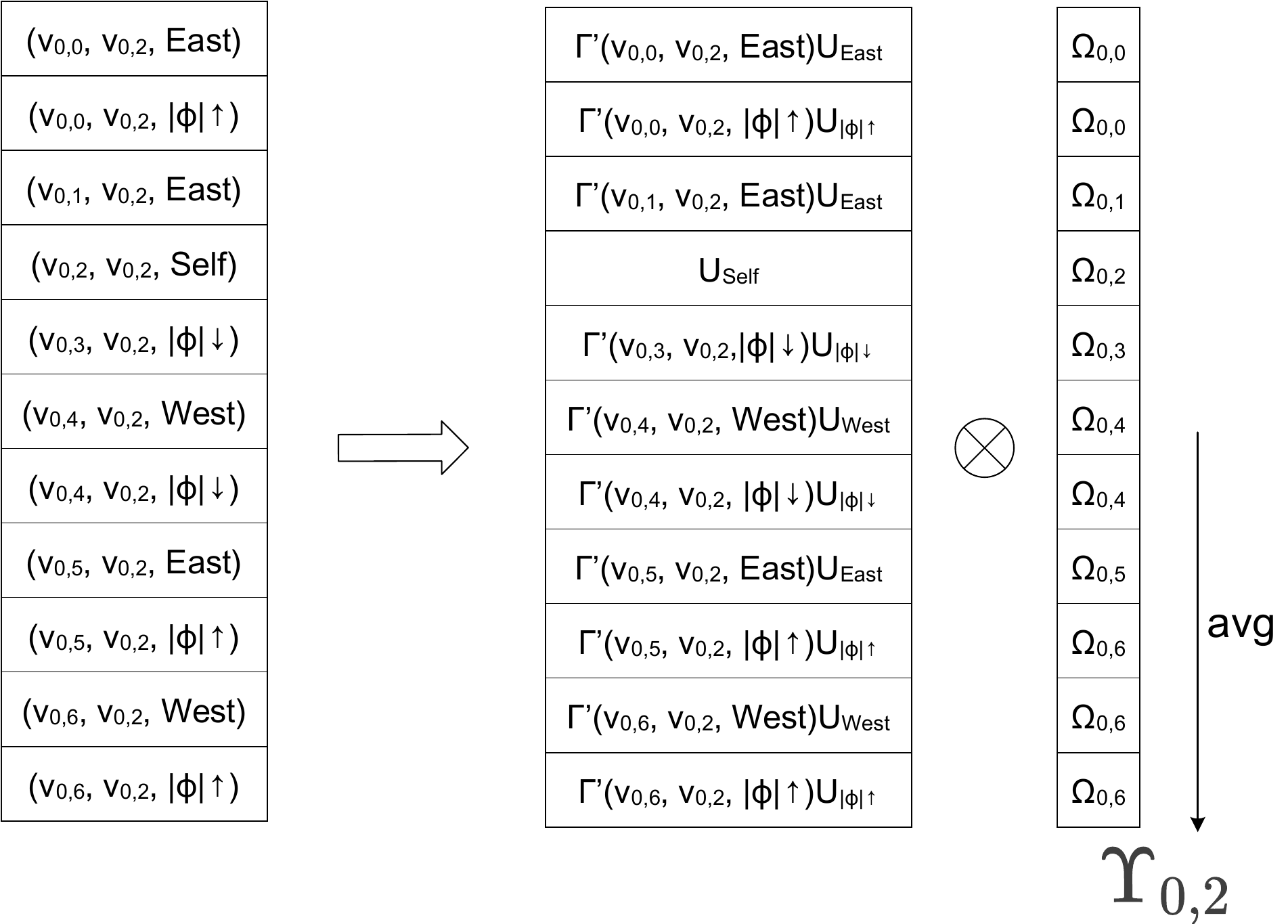}
  \caption{Illustration of graph convolution on $G_0$. 
  The output feature $\Upsilon_{0,2}$ is computed from a weighted sum of input features on node $v_{0,2}$ and its neighbors. 
  Each non-zero edge attribute component in edge pointing to node $v_{0,2}$ corresponds to a weighted matrix, which is the product of the attribute component $\Gamma^\prime(v_{0,j}, v_{0,2}, d)$ and the basis matrix $U_d$.
  }
  \label{fig:conv}
\end{figure}

For ocean ensemble simulations, as input parameters change, severe changes occur only in sub-areas, which means we can represent insensitive regions, where signals slightly change regarding different simulation parameters, with low resolutions to save memory. 





\textbf{Regions' sensitivities.} \quad
We compare simulation outputs with a reference to model regions' sensitivities to the input parameter.
We run $N_1$ simulations for $N_1$ simulation outputs and use the medoid of these outcomes as the reference.
As a learning task, GNN-Surrogate learns the difference between future simulation outputs to the reference, where insensitive regions have low difference values and hence can be represented using low resolutions. 
When running a future simulation, an output residual is defined as the difference between the future simulation output and the reference. 
We calculate the difference between the $N_1$ simulation outputs and the reference for $N_1$ simulation output residuals.

\textbf{Graph hierarchical tree cutting.} \quad
We perform a cutting on our graph hierarchical tree for each simulation output residual, such that computation and memory are reduced to support training GNN-Surrogate given limited computing resources. 
For example, in Figure~\ref{fig:coarsen}(b), nodes above the orange line form a valid cut. 
The goal of the tree cutting is to group graph nodes in contiguous regions that have absolute residual values smaller than a threshold $\epsilon$; those nodes with smaller absolute residual values will be cut and represented by their ancestor where an ancestor represents a group of those nodes.
The selection of $\epsilon$ can be found in the supplementary material. 
For later data reconstruction, each such ancestor is attached with one single residual value, which is the average of the residual values of its leaf descendants.
After the cutting, we call the remaining hierarchical tree $C$, which has a property that, if one tree node is in $C$, all of its ancestors and siblings are also in $C$. 
We determine whether a node is cut or not by examining whether all its leaf descendants have absolute residual values smaller than a threshold $\epsilon$. 
After the remaining tree $C$ is generated, we can use its lowest remaining ancestor as a proxy node to reconstruct the signal of a node being cut. 
We define the proxy node of a cut node $v_{l, j}$ as 
\begin{equation}\nonumber
\psi_C(v_{l, j}) = 
\left\{
\begin{array}{lr}
    v_{l, j} & \textrm{if} \ v_{l, j} \in C, \\
    v_{l^\prime, k} & \textrm{otherwise},
\end{array}
\right.
\end{equation}
where $v_{l^\prime, k} (l^\prime > l)$ is $v_{l,j}$'s lowest ancestor in $C$. 

Our next goal is to integrate multiple cuts of different output residuals into one to build a unified adaptive-resolution representation for future simulation runs. 
For ocean simulation, high-frequency features are usually at specific locations, such as regions near the equators where the eastern equatorial Pacific (EEP) cold tongue happens, making separate cuts similar. 
The guideline to our cut integration is that a node $v_{l,j}$ is cut after the integration only if its parent has the approximating ability for all the residuals. 
If $N_1$ simulations are run to generate $N_1$ residuals for the GHT cut, the time complexity for generating the GHT cut is $O(N_1 \times |V_0|)$. 

After obtaining the integrated GHT cut, a graph $G_l$ is transformed into $G_l^\prime$, where we aggregate nodes and edges. 
For example, in Figure~\ref{fig:coarsen}(c), based on the tree cut shown in Figure~\ref{fig:coarsen}(b), $G_0$ is transformed into $G_0^\prime$, as shown in Figure~\ref{fig:coarsen}(d). 
Specifically, for graphs at each level, we first calculate the reduced node set $V_l^\prime$ by replacing the nodes in $V_l$ with their proxy nodes. 
Second, we generate the new edge set $E_l^\prime$ and edge attribute vectors $\bm{\Gamma_l^\prime}$. 
For two proxy nodes $p_A$ and $p_B$, if there are edges connecting nodes whose proxy are $p_A$ and nodes whose proxy are $p_B$, then $p_A$ and $p_B$ are linked. 
Each new edge attribute vector is the average of edge attribute vectors from edges connecting nodes whose proxy is $p_A$ and nodes whose proxy is $p_B$.
After the graph transformation, a transformed graph hierarchy $\{G_0^\prime, G_1^\prime, \ldots, G_{L-1}^\prime\}$ with an edge attribute vector list $\{\bm{\Gamma_0^\prime}, \bm{\Gamma_1^\prime}, \ldots, \bm{\Gamma_{\bm{L-1}}^\prime\}}$ is generated.
The graph transformation algorithm pseudocode can be found in the supplementary material. 

\subsection{Generation of Training Data with Adaptive Resolutions}
\label{subsection:collect}
The training dataset comprises data pairs of input simulation parameters and the corresponding residuals with adaptive resolutions. 
We have collected $N_1$ ensemble members for training when we build hierarchical tree cuts in Section~\ref{subsection:cut}. 
We further run $N_2$ simulations to generate additional training data. For the new $N_2$ simulations, we save the simulation outputs with adaptive resolutions (without saving full-resolution raw data) to the disk to reduce I/O costs. All the $N_1+N_2$ ensemble members constitute the training dataset. 

\section{Architecture and Operations}
\label{section:network}

We explain the architecture of GNN-Surrogate and related operations in this section. 

\subsection{Architecture}
\label{subsection:architecture}
GNN-Surrogate (denoted as $R$) is an upsampling-convolution generator and is trained to generate outputs close to the ground truth adaptive resolution residuals. 
The network architecture with an upsampling-convolution generator proved effective and computationally efficient ~\cite{ isola2017image, he2019insitunet}.
Here we provide the architecture details.

\begin{figure}
  \centering
  \includegraphics[width=1\linewidth]{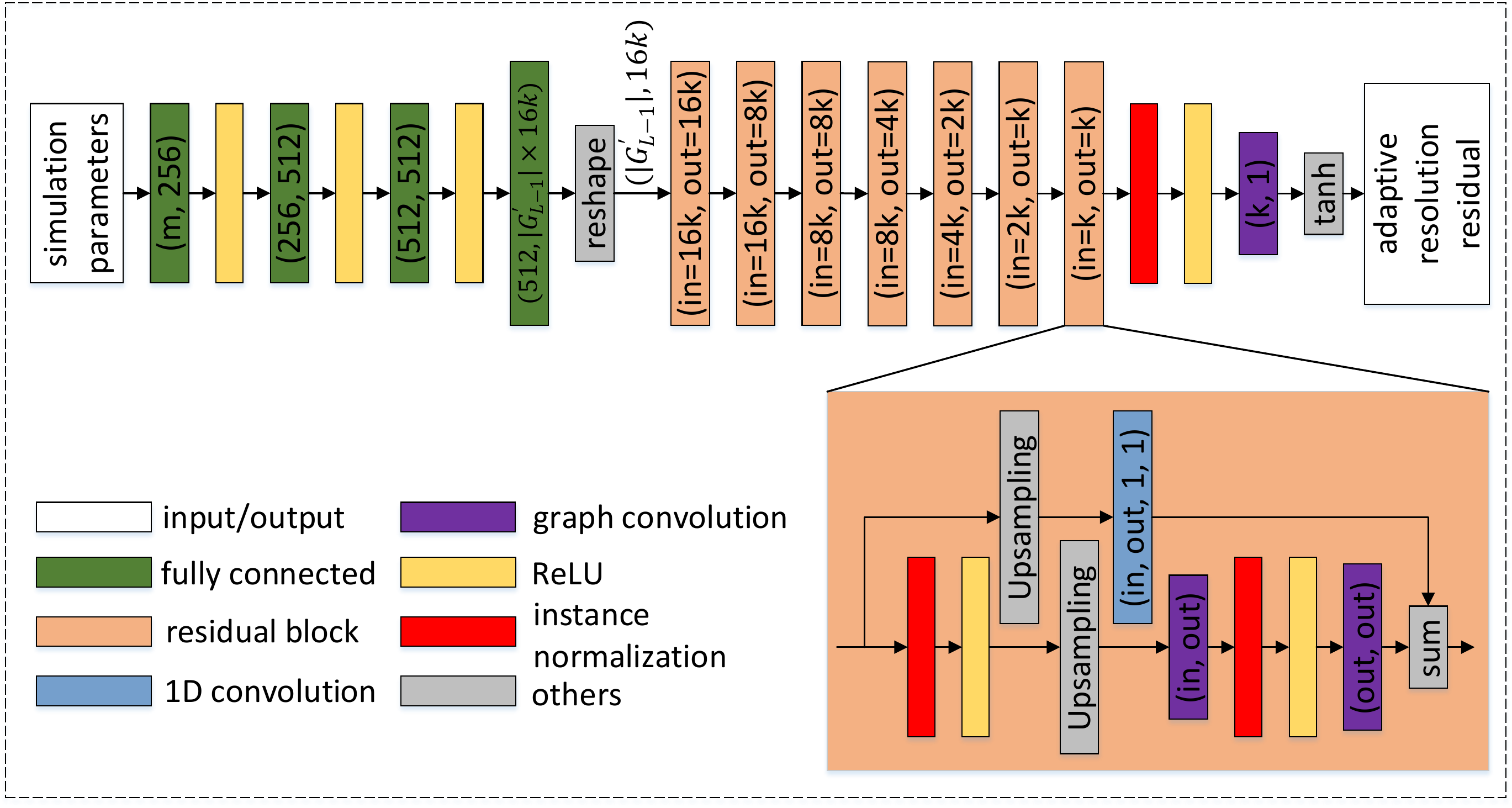}
  \caption{Upsampling-convolution generator architecture of GNN-Surrogate. }
  \label{fig:generator}
\end{figure}
Figure~\ref{fig:generator} shows the architecture of $R$, which takes $P_{sim}$ as input and outputs the adaptive resolution residual data. 
A forward pass in GNN-Surrogate contains three steps. 
First, the input parameter $P_{sim}$ is fed into fully connected layers to generate a latent vector. 
Second, the latent vector is reshaped to a feature map $\Omega_{L-1} \in \mathbb{R}^{|G_{L-1}^\prime| \times 16k}$ on the coarsest graph, where the channel multiplier $k$ is a hyperparameter used to control the network's size.
Third, the subsequent residual blocks perform super-resolution.
A feature map is first upsampled in a residual block and then fed into two graph convolution layers. 
Finally, we add an upsampled original input feature map to the output and send it to the next residual block. 
Note that we use Instance Normalization~\cite{ulyanov2016instance} to speed up the network convergence because Instance Normalization can improve generative models since it can maintain independence between each data instance. 
The rectified linear unit (ReLU)~\cite{nair2010rectified} is used as the activation function in all layers except the last layer. 
The $tanh$ activation function is applied in the last layer to normalize output signals into $[-1,1]$.

\subsection{Graph Convolution}
\label{subsection:convolution}

A convolution operation convolves the input feature map using a convolution layer and passes the result to the next layer. 
Figure~\ref{fig:conv} provides an example of the defined graph convolution.
We define a graph convolution on the constructed graphs as follows. On graph $G_l^\prime$, the convolution operation transforms input feature map $\Omega_l \in \mathbb{R}^{|V_l^\prime| \times c_{in}}$ to an output feature map $\Upsilon_l \in \mathbb{R}^{|V_l^\prime| \times c_{out}}$, where $c_{in}$ and $c_{out}$ are input and output feature maps' channel numbers. 
An output feature $\Upsilon_{l,i} \in \mathbb{R}^{c_{out}}$ is computed from a weighted sum of input features on node $v_{l,i}$ and all of its neighboring nodes. 

We use an edge-conditioned convolution operation formalized as below:
\begin{equation} 
\Upsilon_{l,i} = \dfrac{1}{|\eta(l,i)| + 1} (U_{Self}\Omega_{l,i} + \sum_{(v_{l,j}, v_{l,i}, d) \in \eta(l,i)} \bm{\Gamma_l^\prime}(v_{l,j}, v_{l,i}, d) U_d \Omega_{l,j}),
\label{equation:convolution}
\end{equation}
where $\eta(l,i) = \{ (v_{l,j}, v_{l,i}, d) \ | \ \bm{\Gamma_l^\prime}(v_{l,j}, v_{l,i}, d) > 0\} $
is an edge set containing all the edges pointing to node $v_{l,i}$, $c_{in}$ and $c_{out}$ are channel numbers of the input and output features maps, and $U_{d} \in \mathbb{R}^{c_{out} \times c_{in}} (d \in \mathscr{D})$ and $U_{self} \in \mathbb{R}^{c_{out} \times c_{in}}$ are weighted basis matrices in a convolution layer. 
In each convolution layer, we use different weighted basis matrices conditioned on neighbors at different directions in Equation~\ref{equation:convolution}. 
In our experiments, we find that our model's learning ability increases by exploiting the geographical directional relationship between two edge-connected nodes. 

In our implementation, we first perform linear transformation for the input feature map by dense matrix-vector multiplication, which costs $O(|V_l^\prime|\cdot c_{in}c_{out})$ operations. Then we perform the weighted sum by sparse matrix-vector multiplication, which has a cost of  $O(|E_l^\prime|\cdot c_{out})$ operations. 
We perform matrix multiplication operations in parallel for efficient computations by the support of \texttt{PyTorch Sparse}~\cite{pytorch_sparse} library.

\subsection{Graph Upsampling}
\label{subsection:upsample}
This section defines graph upsampling to transform feature maps between graphs of different resolution levels. 

We upsample a graph of level $m$ to a graph of level $l$, where $m > l$. 
The value of a node at level $l$ equals the ancestor at level $m$. 
Given a feature map $\Omega_m \in \mathbb{R}^{|V_m^\prime| \times c}$ on graph $G_m^\prime$, the graph upsampling outputs $\Upsilon_l \in \mathbb{R}^{|V_l^\prime| \times c}$ on graph $G_l^\prime$, such that 
\begin{equation}
\Upsilon_{l, i} = \Omega_{m, j}: v_{m, j}\in V_m^\prime \ \textrm{is the ancestor of} \ v_{l, i} \in V_l^\prime \ \textrm{in the GHT},
\end{equation}
where $c$ is the channel number of the feature map. 

Our graph upsampling operations can be implemented by multiplying an input feature map by a sparse matrix that with $|V_l^\prime| \cdot c$ items.
Upsampling operations on the whole input feature map costs $O(|V_l^\prime| \cdot c)$ operations. 
The same as Section~\ref{subsection:convolution}, 
we perform matrix multiplication operations in parallel for efficient computations by the support of \texttt{PyTorch Sparse}~\cite{pytorch_sparse} library.

\subsection{Training Process}
\label{subsection:network}

In the training process, we update the network parameters in GNN-Surrogate using gradient descent. 
We also explain other used methods during training as follows.
 
\textbf{Loss Function} \quad
During training, we iteratively update parameters in GNN-Surrogate to minimize a loss function. 
Our loss function is a $L_1$ loss:
\begin{equation}
L_1 = 
\dfrac{1}{b}\sum_{i=0}^{b-1}
||\hat{S}_{ar,i} - S_{ar,i}||_1,
\end{equation}
where $b$ is the batch size, $S_{ar,0:b-1}$ and $\hat{S}_{ar,0:b-1}$ are ground truth and generated adaptive resolution residual.
The loss can guide $R$ to produce results with high accuracy. 

\textbf{Training Techniques} \quad
We improve the training stability and efficiency using two additional techniques. 
First, We apply spectral normalization~\cite{miyato2018spectral} to stabilize our GNN-Surrogate training. 
Second, training a deep generative model for large-scale data is memory costly and slow.
Thus, we train GNN-Surrogate with mixed precision~\cite{micikevicius2018mixed}, which can reduce the memory cost of training and the time it takes to train with minimal impact. 
The details of our training techniques can be found in the supplementary material.

\subsection{Inference Process and Sensitivity Analysis}
\label{subsection:sensitivity}

In the inference stage, a new simulation input parameter setting is fed into GNN-Surrogate. 
After a forward pass, we obtain an adaptive resolution residual. 
We convert the adaptive resolution residual to a residual with the full resolution by nearest neighbor sampling. 
Finally, we get the predicted simulation output by adding the reference onto the residual.   

Inspired by previous works~\cite{berger2018generative, he2019insitunet}, we exploit GNN-Surrogate's differentiability to perform sensitivity analysis on simulation parameters. 
We aggregate the predicted field (e.g., L1 norm of the data values) to obtain a scalar value and compute the derivative of that scalar value with respect to one selected parameter. 
The absolute value of the derivative can be used as an indicator for one parameter's sensitivity since it reflects how the field changes as the input simulation parameter change. 
While analyzing one selected parameter, we fix the other parameters and perform uniform sampling in the selected parameter's value range. 
For each sample point, forward propagation and backward propagation are conducted to obtain a sensitivity value. 
We use a line chart to visualize the list of sensitivity values. 

\section{Results}
\label{section:result}
The evaluation for our proposed GNN-Surrogate is broken up into three sections: (1) implementation details and performance  (Section~\ref{subsection:imple}); (2) quantitative and qualitative analyses comparing our approach with baseline approaches (Section~\ref{subsection:comparison}); (3) parameter space exploration case studies and further analysis (Section~\ref{subsection:exploration}).

\subsection{Ensemble Simulation Running Settings}
\label{subsection:ensemble}
MPAS-Ocean is designed for the simulation of the global ocean system. Based on the scientist's suggestion, we studied four parameters: 
the amplitude of the ocean surface wind stress ($BwsA \in [0.0, 5.0]$), the magnitude of the Gent McWilliams mesoscale eddy parameterization ($GM \in [600.0, 1500.0]$), the critical bulk Richardson number (used to determine the strength of vertical mixing) ($CbrN \in [0.25, 1.00]$), and horizontal viscosity ($HV \in [100.0, 300.0]$). 
We randomly sampled 100 parameter settings from the parameter space and randomly picked 70 for training and 30 for testing. 
15-model-day ocean simulations were conducted with each parameter setting, and unstructured grids with EC60to30 resolution representing the ocean temperature were generated. 
An MPAS-Ocean mesh structure contains 60 horizontal layers, and each horizontal layer consists of 235,160 Voronoi cells.   
One generated ensemble member takes 1.00 GB space with a temperature field of size 107.65 MB within it. 
Among the 70 ensemble members in the training dataset, 16 were used for generating the hierarchy cutting policy described in Section~\ref{subsection:cut}. 

\subsection{Implementation and Performance}
\label{subsection:imple}

\textbf{Graph preparation, including graph hierarchy generation and graph transformation} \quad 
The graph preparation was implemented in C++, and the graph is represented with Eigen SparseMatrix. 
  
Our graph preparation was computed on an Intel Xeon E5-2680 CPU. 
The graph construction and graph coarsening took 25.9 minutes. The graph transformation took 6.18 minutes. 
The graph preparation is only related to the mesh structure and independent of simulation outputs, so only one graph preparation was performed. 

\textbf{Simulation runs and data collection}  \quad 
The simulations were conducted on a supercomputer with 648 computation nodes. Each node contains an Intel Xeon E5-2680 CPU with 28 cores and 128 GB memory.
We used 128 processes for our simulation, and it took 49.6 minutes per simulation.
After the simulation, the size of an output adaptive resolution temperature field varies given different GHT cut thresholds and is reported in the supplementary material. 

\textbf{GNN-Surrogate training and inference}  \quad
GNN-Surrogate was implemented in PyTorch~\cite{paszke2019pytorch}. 
The training and testing of GNN-Surrogate were on one NVIDIA Volta V100 GPU 16GB. 
We fixed the GNN-Surrogate training time to $36$ hours. 
After training, a single forward pass through the trained  GNN-Surrogate takes less than 2 seconds, instead of the roughly 50 minutes it would take to directly compute the simulation output on a large cluster of CPUs.
  
\subsection{Comparison with Baseline Approaches}
\label{subsection:comparison}

\subsubsection{Evaluation Metrics}
\textbf{Data-level metrics} \quad
GNN-Surrogate allows scientists to use any visual mapping parameters of interest to the reconstructed simulation output, so it is necessary to evaluate the predicted simulation output's quality at the data level.
We used peak signal-to-noise ratio (PSNR) to measure the grid-level difference, and normalized maximum difference (MD) for the error bound. 

\textbf{Geometry-level metrics}  \quad
The isothermal layer (ITL) depth reflects the local ocean temperature and spatial variability from a geometric perspective. 
In this work, we compute specific ITL depths and calculated measures of overlap (Jaccard coefficient, JC) and the mean surface distance of intersection regions to evaluate the quality of ITL depths. 

\textbf{Image-level metrics}  \quad
At the image level, horizontal and vertical cross-section images were rendered. 
We fixed the depth, latitude, or longitude and used the Kindlmann colormap~\cite{kindlmann2002face} to establish a correspondence between the color and the ocean temperature.
Structural similarity index measure (SSIM) and earth mover’s distance (EMD) between color histograms~\cite{berger2018generative, he2019insitunet} were used to quantify the structural and distributional similarity between two rendered images.

We evaluated both global and specific regions of interest (ROI). 
For ROI evaluation, we extracted a region within $160^\circ W$ to $80^\circ E$, $26^\circ S$ to $26^\circ N$, and sea level to a depth of 200 meters, which is the location of the eastern equatorial Pacific cold tongue. 
At the image level evaluation, the resolution of images depends on its target region, $1024 \times 512$ for images of a global region, and $420 \times 180$ for images of an ROI.

\subsubsection{Baselines}
\label{subsection:baseline}
The baseline methods we chose to compare with are inverse distance weighting (IDW) interpolation for comparison on all analyses, radial basis function (RBF) interpolation for comparison of data-based analyses, and the image-based method InSituNet~\cite{he2019insitunet} for comparison of image-based analyses. 
The reasons we selected IDW interpolation as the baseline method are twofold. 
First, IDW is one of the most commonly used interpolation methods for scientific data analysis~\cite{chen2012estimation}.
Second, IDW interpolation has a low computation cost and is straightforward to interpret~\cite{lu2008adaptive}.
Radial basis function (RBF) interpolation is a more complicated interpolation method and is also widely used for scientific data~\cite{wild2008orbit}. 
We used Gaussian distribution as the radial basis function and used backward propagation for the optimization.
We selected InSituNet for comparison of image-based analyses because it is the state-of-the-art image-based method. 
For the IDW interpolation method, we searched the training dataset, sampled $g$ data instances whose parameter settings have the minimum Manhattan distance to the test data, and applied the weighted sum. 
Values of $g$ from 1 to 5 were evaluated, and we present the results for $g=3$ since it balances the metrics in three levels (more results are in the supplementary material). 
The CPU time required to get an interpolated result is around 1.3 seconds, which is a little smaller than a single forward pass time through the trained GNN-Surrogate.
For InSituNet, since it is an image-based method, it does not support data level or geometric level comparison, nor does it support multiple visual mappings. 
Thus, we only rendered the horizontal cross-sections with a fixed color mapping as the target output of InSituNet.  
The corresponding simulation parameters and depth were incorporated as the input parameters.  

\subsubsection{Quantitative and Qualitative Analysis Results}

\begin{figure}
  \centering
  \includegraphics[width=1\linewidth]{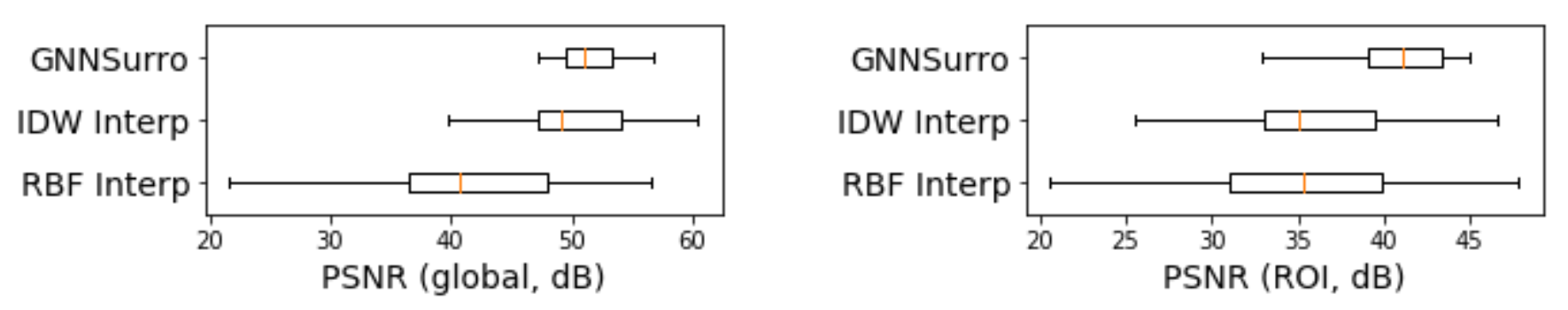}
  \caption{The box plot showing the PSNR deviation on global and ROI temperature maps from 30 different testing ensemble members. }
  \label{fig:box}
\end{figure}

\begin{figure*}
  \centering
  \includegraphics[width=\linewidth]{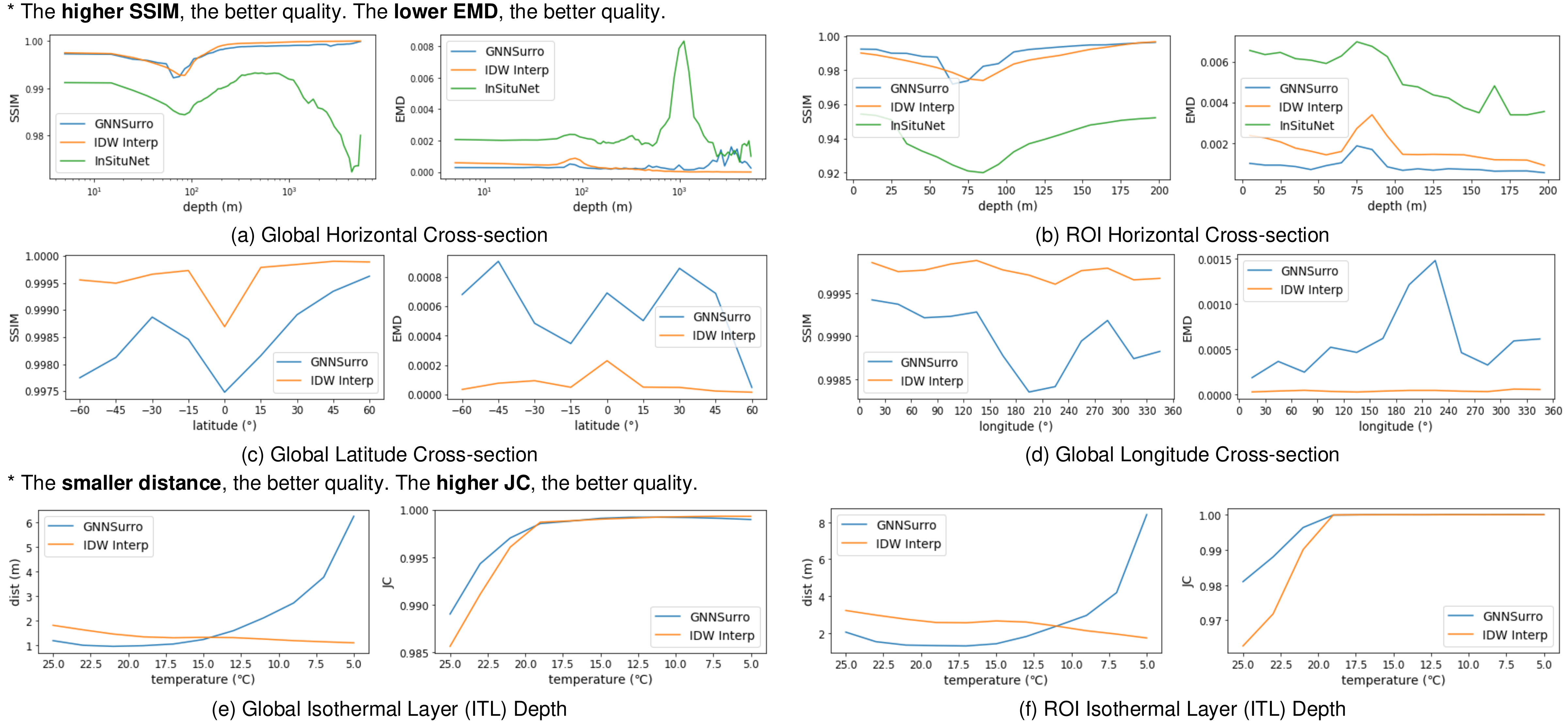}
  \caption{(a-b) SSIM and EMD for temperature horizontal cross-sections at different depths. 
  (c) SSIM and EMD for vertical cross-sections at different latitudes. 
  (d) SSIM and EMD for vertical cross-sections at different longitudes. 
  (e-f) Average distance and Jaccard Coefficient (JC) for the isothermal layer (ITL) depth maps with different temperature isovalues. }
	\label{fig:plot}
\end{figure*}

\begin{figure*}
  \centering
  \includegraphics[width=1\linewidth]{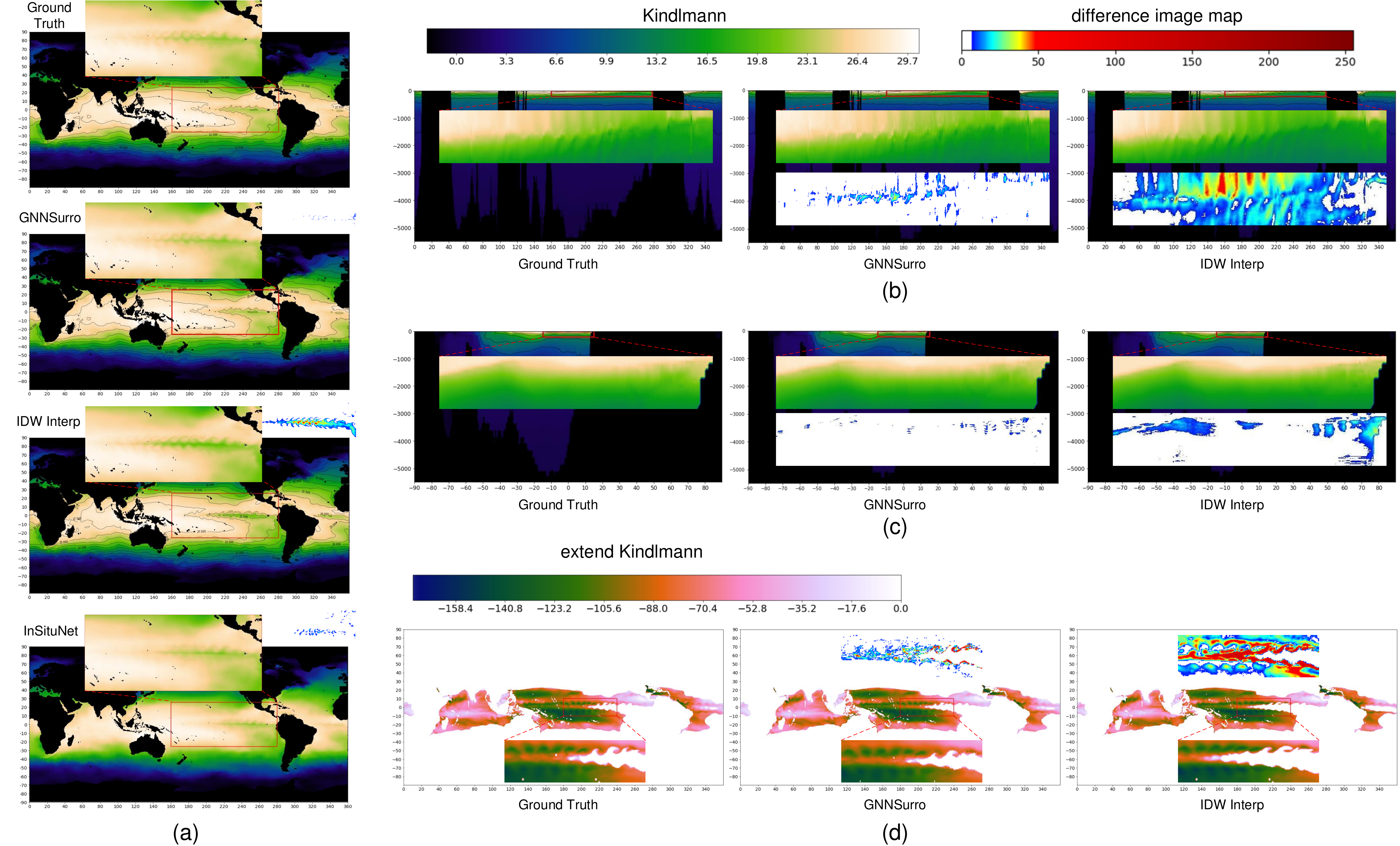}
  \caption{ (a) Comparison of the sea level temperature maps generated using GNN-Surrogate, IDW interpolation, and InSituNet with the ground truth maps.
  Comparison of the vertical cross-sections at (b) the equator (c) $75^\circ E$ generated using GNN-Surrogate and IDW interpolation with the ground truth cross-sections. 
  (d) Comparison of the isothermal layer (ITL) depth maps with temperature isovalue $25^\circ C$ generated using GNN-Surrogate and IDW interpolation with the ground truth maps. }
  \label{fig:map}
\end{figure*}

The evaluation results are presented from three perspectives: (1) data-level analysis using Table~\ref{table:psnr} and Figure~\ref{fig:box}, (2) geometry-level analysis using  Figure~\ref{fig:map}(d) and Figure~\ref{fig:plot}(e, f), and (3) image-level analysis using Figure~\ref{fig:map}(a-c) and Figure~\ref{fig:plot}(a-d).

\begin{table} \footnotesize
\caption{Quantitative comparison of the output predicted with GNN-Surrogate, radial basis function (RBF) interpolation, and inverse distance weighting (IDW) interpolation. }
    \centering
    \begin{tabular}{l|c|c|c} 
           & GNN-Surrogate & IDW Interp & RBF Interp \\  \hline
         PSNR (global, dB) & \textbf{50.7, 2.52} & 47.7, 5.72 & 32.43, 8.91  \\  \hline
         MD (global) & 0.1965, 0.0415 & 0.1721, 0.0562 & \textbf{0.1397, 0.0398}  \\ \hline
         PSNR (ROI, dB) & \textbf{39.5, 3.06} & 33.6, 6.03 & 27.43, 7.27 \\  \hline
         MD (ROI) & 0.1774, \textbf{0.0332} & 0.1673, 0.0672 & \textbf{0.1573}, 0.0605  \\  \hline
         params (GB) & \textbf{2.18} & 7.18 & 10.27    \\ 
    \end{tabular}
    \label{table:psnr}
\end{table}

\textbf{Data-level analysis} \quad
In Table~\ref{table:psnr}, we quantitatively compared GNN-Surrogate results against interpolations at the data level (PSNR and MD) both globally and in the ROI. 
We found that globally, the GNN-Surrogate produces higher PSNR than interpolations. 
GNN-Surrogate has a worse normalized maximum difference than interpolations, which is explainable because our loss function does not constrain the error bound. 
Both interpolations and GNN-Surrogate have lower PSNR in the ROI than global PSNR because the ROI contains complex phenomena. 
GNN-Surrogate achieves a relatively higher PSNR than interpolation methods for the ROI compared to the entire domain. 
GNN-Surrogate is more stable than interpolations since it usually has a smaller standard variance. 
To illustrate that, we give PSNR box plots in Figure~\ref{fig:box}. 
Although sometimes interpolations may give better prediction results than GNN-Surrogate, they have worse lower quartiles and medians, making them less trustworthy. 

\textbf{Geometry-Level analysis} \quad
We sampled temperature isovalues from $25^{\circ}C$ to $5^{\circ}C$ and computed the ITL depth. 
In Figure~\ref{fig:plot} (e, f) the quantitative results are shown.
We arrange temperature from large to small in the figure because ocean temperature usually monotonically decreases as the depth increases, so a larger temperature isovalue leads to a smaller depth.
GNN-Surrogate has a smaller mean surface distance than IDW interpolation when the temperature is higher than $15^{\circ}C$. 
The distance is always smaller than ten meters, which is good considering that the ocean's mean depth is 5.5km. 
Note that as temperature decreases, there is a rapid upward trend for the GNN-Surrogate's mean surface distance, while for IDW interpolation, the mean surface distance gradually decreases.
We believe the rapid upward trend is because of the lower vertical resolution in the deeper ocean. 
Meanwhile, the temperature is not sensitive to any input parameters in this region, so IDW interpolation can still do a good job. 
For the surface overlap, evaluated by Jaccard Coefficient, GNN-Surrogate is slightly better than IDW interpolation until they both become almost perfect when the temperature goes below $19^{\circ}C$.

In Figure~\ref{fig:map}(d), we compared rendering results of the $25^{\circ}C$ ITL depth generated by IDW interpolation and GNN-Surrogate using the Extended Kindlmann colormap~\cite{moreland2016we}. 
GNN-Surrogate generates a depth map close to the ground truth. 
In the depth map generated by IDW interpolation, there is a large gap across the equator in the eastern Pacific. 
The depth map rendering does not always give enough information. 
For example, we do not know why there is a strange gap. 

\textbf{Image-Level analysis} \quad
Apart from the rendering using the Kindlmann colormap, inspired by some previous work~\cite{han2019tsr, han2020v2v}, difference images are given to display the noticeable pixel differences (with $\triangle \geq 6.0$ in the CIELUV color space). 

We rendered different horizontal cross-sections from the sea surface to the seabed. 
First, in Figure~\ref{fig:plot}(a), GNN-Surrogate outperforms InSituNet, reflected from higher SSIM and lower EMD.
Second, although GNN-Surrogate performs well, it is slightly worse than IDW interpolation for the global region at some depths. 
We think this is because the temperature is not sensitive to any input parameters not only in the deep ocean but also at many locations close to the sea level. 
To illustrate that, we calculate SSIM and EMD for ocean maps in the ROI and find that GNN-Surrogate achieves higher SSIM at most depths and always has lower EMD, as seen in~\ref{fig:plot}(b).  
Figure~\ref{fig:map}(a) shows the sea level rendering results. 
GNN-Surrogate accurately reflects the ocean temperature, while IDW interpolation incorrectly predicts temperatures of the equatorial cold tongue (i.e., predicted temperatures are much lower than the ground truth), which explains why we find a large gap in the $25^{\circ}C$ ITL depth.
InSituNet has two major limitations. 
First, since it directly predicts the image, we cannot easily add isotherms to aid visualization. 
Moreover, it cannot preserve interesting features, such as the equatorial cold tongue, well. 

We give the vertical cross-sections of the temperature field as well. 
In figure~\ref{fig:plot}(c, d), we observe that IDW interpolation generates images with higher SSIM and lower EMD. 
We claim that ocean temperature usually monotonically decreases as depth increases, which makes the vertical cross-sections featureless. 
Despite that, in some particular regions, GNN-Surrogate can be more trustworthy. 
For example, in Figure~\ref{fig:map}(b), the vertical cross-section from the equator, we can see IDW interpolation predicts the temperature to be lower than the ground truth in the Pacific near the sea level.
In Figure~\ref{fig:map}(c), a vertical cross-section from $75^\circ E$, we find that IDW interpolation predicts the temperature to be lower near the continent at around $10^\circ N$. 
GNN-Surrogate performs better on these two examples.  

\subsection{Case Study: Parameter Space Exploration}
\label{subsection:exploration}
In this section, a scientist (one of our coauthors) who has 15 years of ocean science experience exploited GNN-Surrogate to do a parameter space exploration on an MPAS-Ocean simulation forced by an annual averaged wind stress and restoring of surface temperature and salinity to climatology.  

\begin{figure}
  \centering
  \includegraphics[width=1\linewidth]{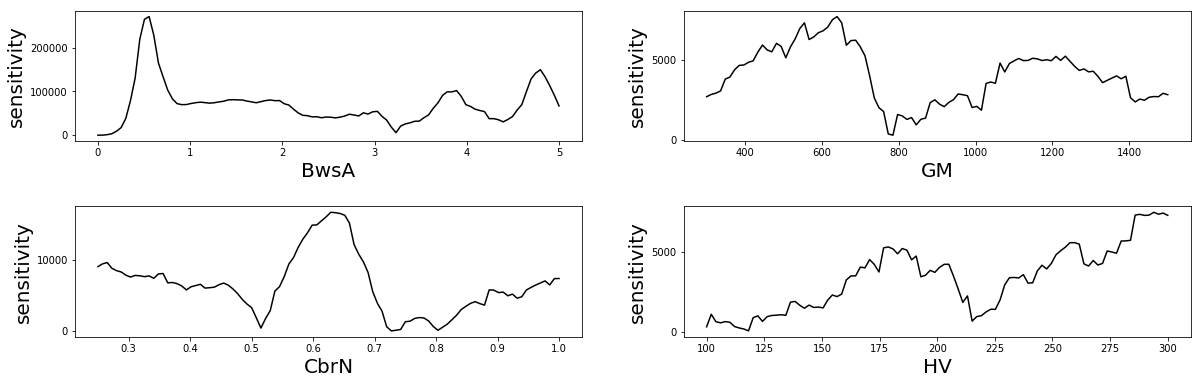}
  \caption{ The sensitivity line graph visualization of different simulation parameters.  }
  \label{fig:sensitivity}
\end{figure}

First, he probed the sensitivity to different parameters. 
Under the parameter setting $BwsA = 2.5, GM = 900.0, CbrN = 0.625, HV =  200.0$, he computed the sensitivity of each variable in turn, as suggested in Section~\ref{subsection:sensitivity}. 
The results are shown in Figure~\ref{fig:sensitivity}. 
Considering the scale of the four plots, he ranked variable sensitivity: $BwsA > CrbN > GM > HV$. 

\begin{figure}
  \centering
  \includegraphics[width=1\linewidth]{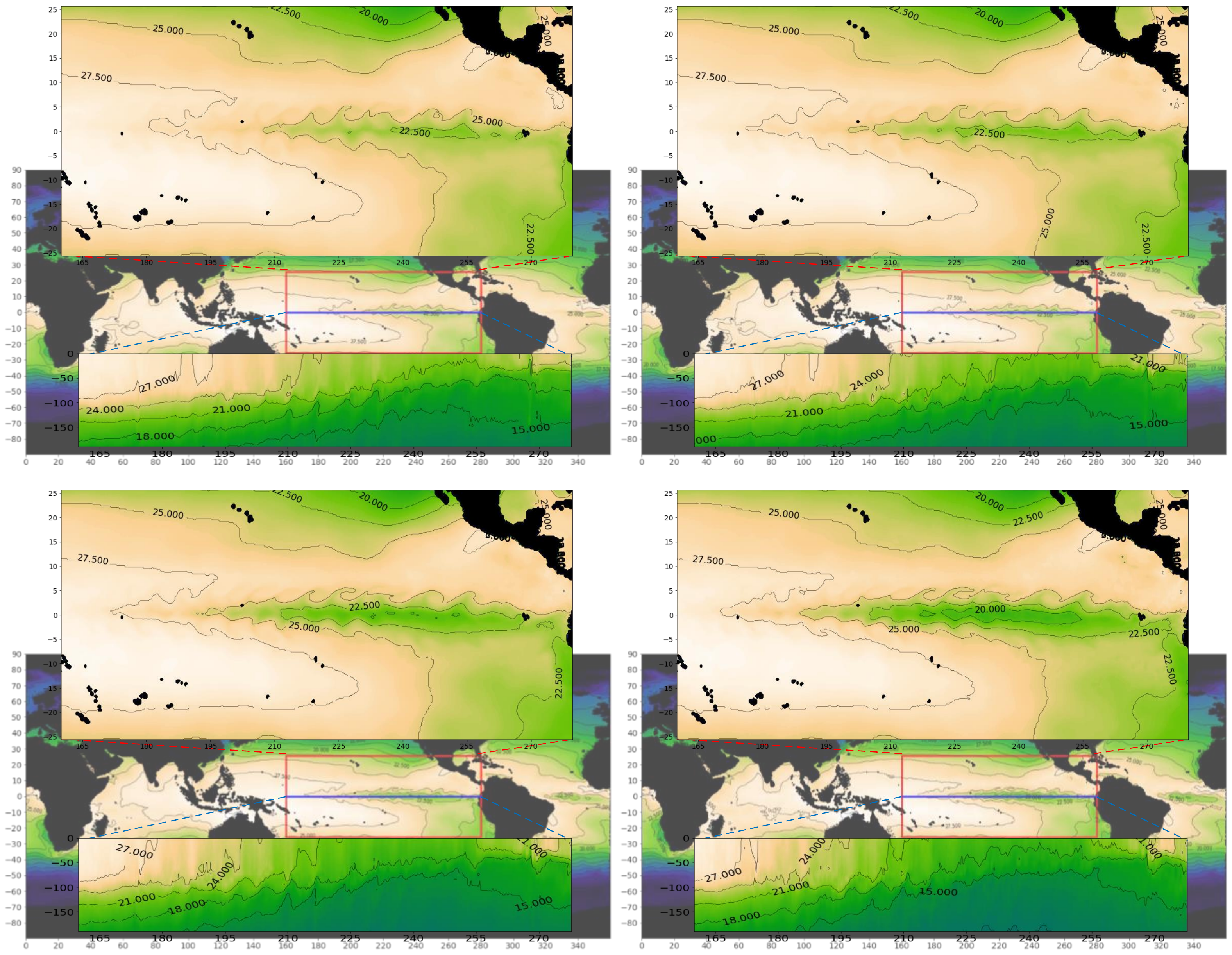}
  \caption{Comparison of the sea level temperature map and equator vertical cross-sections using different $BwsA$ values to see the effect of the amplitude wind stress. 
  As the wind stress becomes strong ($BwsA$ value becomes high), the equatorial cold tongue in the eastern Pacific is significantly enhanced, and more cold water in the deeper ocean is upwelled to the surface.}
  \label{fig:case_map}
\end{figure}

Next, he examined how $BwsA$, the most sensitive parameter, affects the ocean temperature. 
He fixed $GM = 900.0, CbrN = 0.625, HV =  200.0$, and sampled $BwsA$ from $\{0.0, 1.0, 3.0, 5.0\}$. 
As shown in Figure~\ref{fig:case_map}, he first visualized the sea surface temperature map.
Based on the observation, as he scaled up the amplitude wind stress, the equatorial cold tongue in the eastern Pacific is significantly enhanced. 
This phenomenon is physically expected as the trade winds, which blow from east to west, push the warm ocean surface water toward the western Pacific, exposing the cooler surface waters below.  
Further, the increased wind leads to the upwelling of colder subsurface water along the equator due to the Coriolis force.  
These two processes lead to a stronger equatorial cold tongue.
To verify this point, he rendered the east Pacific equatorial cross-section from the sea level to a depth of 200 meters, as shown in Figure~\ref{fig:case_map}. 
As the wind stress becomes strong, more cold water in the deeper ocean is upwelled to the surface. 

\section{Discussion and Limitations}
\label{section:discussion}

We compare our method with an image-based surrogate model InSituNet~\cite{he2019insitunet} and list two limitations of our method in this section. 

We addressed all the three limitations reported in InSituNet: (1) insufficient flexibility when exploring arbitrary visual mapping parameters; (2) low accuracy of predicted images; (3) low resolution of predicted images.
For the first limitation of InSituNet, instead of considering the huge joint space of all the simulation and visualization parameters, GNN-Surrogate predicts the simulation output first. 
Using adaptive resolutions for MPAS-Ocean data, our training dataset size is 3.13GB, smaller than InSituNet's training data requirement (3.46GB). 
After simulation outputs are predicted, various existing visualization algorithms can be used for rendering. 
For the second and third limitations of InSituNet, we exploit the fact that all the ensemble members share the same mesh structure, which does not need to be learned. 
Meanwhile, GNN-Surrogate builds adaptive resolutions for simulation outputs, which significantly reduces the training difficulty. 
For the training dataset generation, compared with InSituNet, GNN-Surrogate requires fewer ensemble runs (70 versus 270) and learns more input parameters (4 versus 1). 
With the high quality of the predicted simulation output, accuracy and resolution of rendered images are improved significantly, which were validated by our experiments. 

GNN-Surrogate has two limitations that we want to research for the solution in the future: 
(1) long offline training time and (2) limited sensitivity estimation before the model is trained. 
First, the offline training time of GNN-Surrogate takes $36$ hours.
Thanks to the advancement of high-performance machine learning, we plan to train our GNN-Surrogate with multiple GPUs using the data-parallel technique provided by PyTorch~\cite{paszke2019pytorch} to speed up the training in the future. 
Second, to estimate the sensitivity of regions for building adaptive resolutions, we only select one reference by selecting the medoid of the first $N_1$ simulation outputs. 
Considering the complexity of the simulation parameter space, one reference may not always be representative of modeling the sensitivities of regions against other ensemble members. 
In the future, we plan to compute region's sensitivities from multiple simulation outputs as references for a more accurate sensitivity estimation to model adaptive resolutions.  

\section{Conclusions and Future Work}
\label{section:conclusion}
In this work, we propose GNN-Surrogate, a deep surrogate model to support analysis and visualization of ocean ensemble simulations.
GNN-Surrogate is based on a graph neural network for learning on data represented with unstructured grids, and adaptive data resolution is applied for GPU memory efficiency.
In the inference stage, GNN-Surrogate is fed with input parameters, and the output can be converted to the full resolution by simple post-processing. 
Scientists can apply existing visualization algorithms to the reconstructed data and conduct a thorough analysis. 
We render horizontal and vertical cross-sections and ITL depth maps and give comprehensive quantitative and qualitative evaluations to demonstrate the effectiveness and efficiency of GNN-Surrogate.

\textbf{Future work} \quad
In Section~\ref{subsection:sensitivity}, we used nearest neighbor sampling to convert the predicted adaptive resolution result to full resolution.
Although it worked well in our experiments, in the future, we plan to research on neural network-based super-resolution method to further improve GNN-Surrogate's quality. 
In Section~\ref{subsection:baseline}, we chose a computationally cheap and straightforward inverse distance weighting interpolation and a more advanced radial basis function interpolation as baseline methods.
Some other techniques like Kriging can also be considered. 
However, Kriging is a geostatistical approach that requires calculating corresponding interpolation coefficients from sampled temperature values for each spatial location. 
Considering that the number of spatial grid points in one ensemble member is very high ($10^7$), Kriging is not appropriate as the comparison baseline due to a high computation cost.
In the future, we would like to explore interpolation techniques that have higher performance and competitive computational cost.
We would also like to extend GNN-Surrogate to time-varying data by employing 4D convolutions, time-space graphs and trees. 
Outside ocean simulations, GNN-Surrogate can also be useful for other datasets. 
However, in that case, our adaptive data resolution method may not work well since different ensemble members may not have features at the same location.
We consider using another model to learn the graph transformation strategy and guide our adaptive data resolution to overcome this challenge. 

\ifCLASSOPTIONcompsoc
  \section*{Acknowledgments}
\else
  \section*{Acknowledgment}
\fi

This work is supported in part by the National Science Foundation Division of Information and Intelligent Systems-1955764, the National Science Foundation Office of Advanced Cyberinfrastructure-2112606, U.S. Department of Energy Los Alamos National Laboratory contract 47145, and UT-Battelle LLC contract 4000159447 program manager Margaret Lentz. 
This work is also supported in part by the SciDAC program funded by the U.S. Department of Energy, Office of Science, Advanced Scientific Computing Research. 
This research used resources of the Argonne Leadership Computing Facility, which is a DOE Office of Science User Facility supported under Contract DE-AC02-06CH11357.

\ifCLASSOPTIONcaptionsoff
  \newpage
\fi



\bibliographystyle{IEEEtran}
\bibliography{template}
%


%

\begin{IEEEbiography}[{\includegraphics[width=1in,height=1.25in,clip,keepaspectratio]{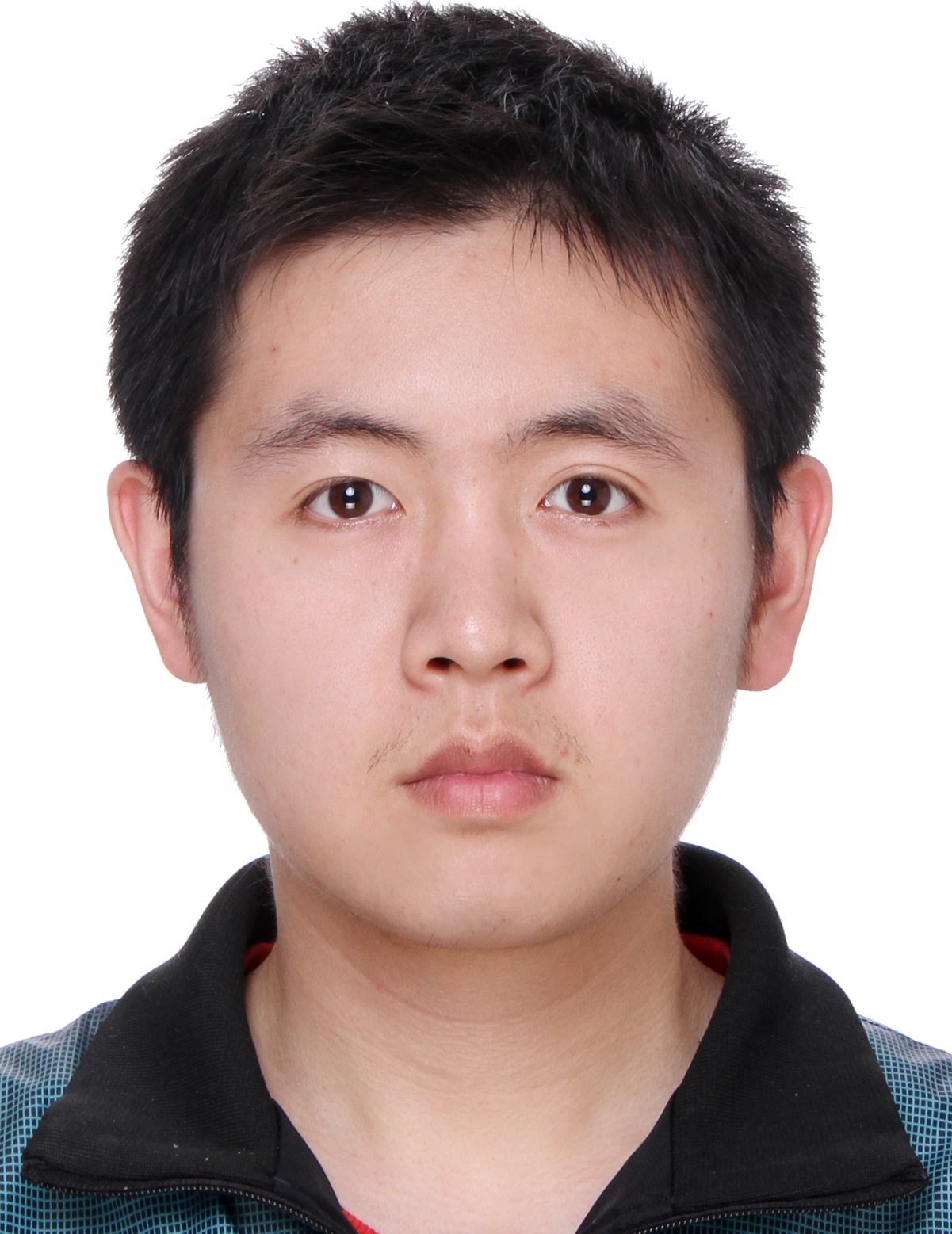}}]{Neng Shi}
is a Ph.D. student in the Department of Computer Science and Engineering at the Ohio State University.  
He received his B.S. degree in Geographic Information Science from Zhejiang University in 2018.
His research interest include large-scale scientific data visualization, ensemble simulation data visualization, and machine learning for scientific visualization. 
\end{IEEEbiography}

\begin{IEEEbiography}[{\includegraphics[width=1in,height=1.25in,clip,keepaspectratio]{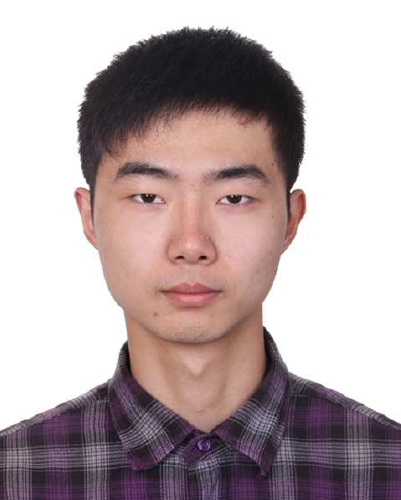}}]{Jiayi Xu}
is a research scientist at Meta AI. His research interests include high-performance data analysis, visualization, and machine learning. He is the recipient of the best paper award in the 14th IEEE Pacific Visualization Symposium. He received his Ph.D. degree in computer science and engineering from The Ohio State University in 2021 and his B.E. degree in computer science and technology from Chu Kochen Honors College of Zhejiang University in 2014. 
\end{IEEEbiography}

\begin{IEEEbiography}[{\includegraphics[width=1in,height=1.25in,clip,keepaspectratio]{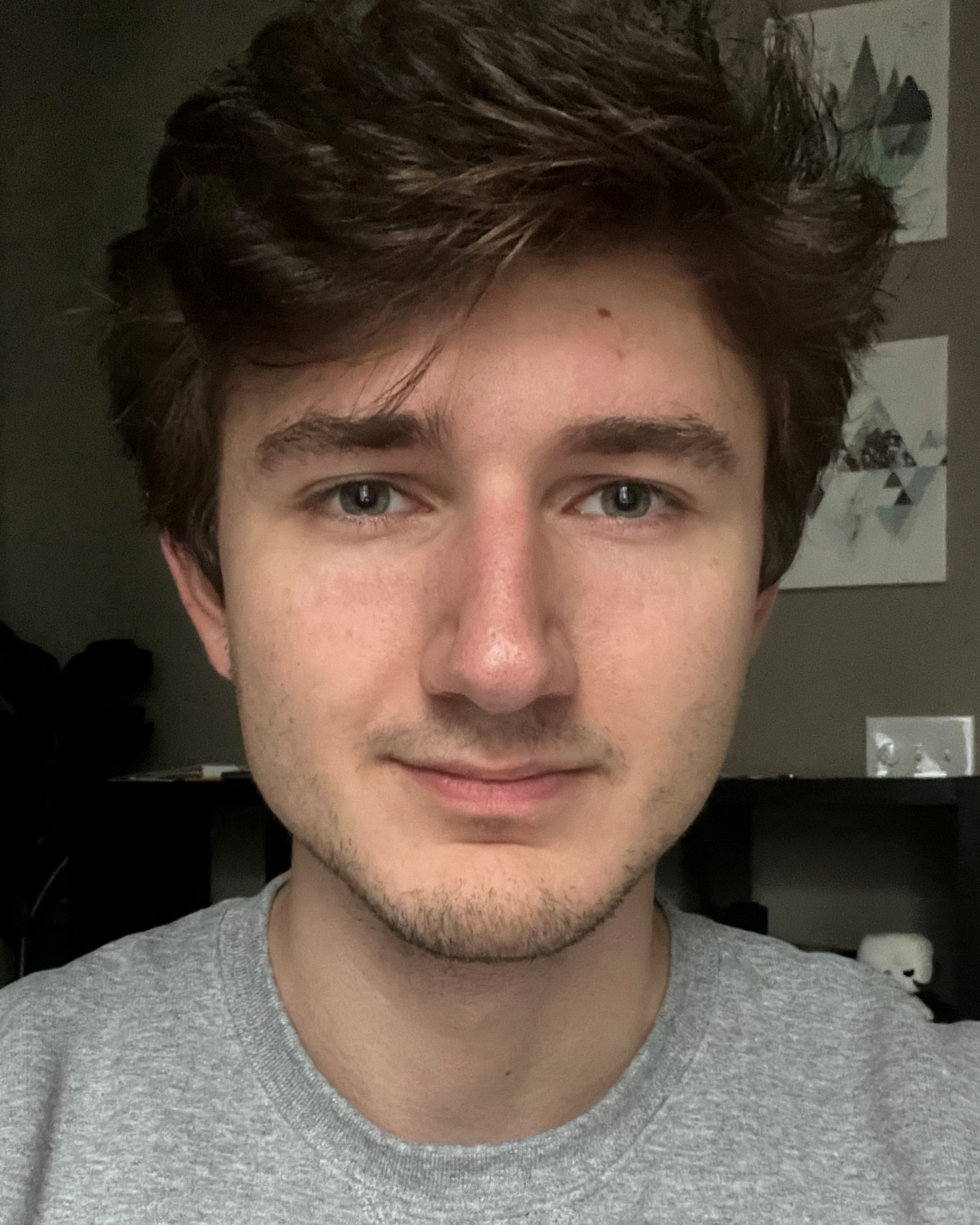}}]{Skylar W. Wurster} is a 3rd year Ph.D. student studying under Professor Han-Wei Shen as part of his GRAVITY research group at The Ohio State University in Columbus, Ohio. He also collaborates with mentors Hanqi Guo and Tom Peterka at Argonne National Lab in Lemont, Illinois. His research interests span deep learning, scientific data visualization, and computer games/graphics.
 
\end{IEEEbiography}

\begin{IEEEbiography}[{\includegraphics[width=1in,height=1.25in,clip,keepaspectratio]{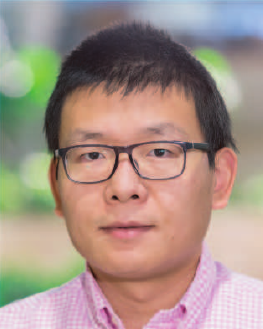}}]{Hanqi Guo}
is a computer scientist at Argonne National Laboratory, scientist at the University of Chicago Consortium for Advanced Science and Engineering (CASE), and fellow of the Northwestern Argonne Institute for Science and Engineering (NAISE).  His research interests include data analysis, visualization, and machine learning for scientific data. He is the recipient of multiple best paper awards in IEEE VIS, IEEE PacificVis, and ChinaVis.  He received his Ph.D. degree in computer science from Peking University in 2014 and his B.S. degree in mathematics and applied mathematics from Beijing University of Posts and Telecommunications in 2009.
\end{IEEEbiography}

\begin{IEEEbiography}[{\includegraphics[width=1in,height=1.25in,clip,keepaspectratio]{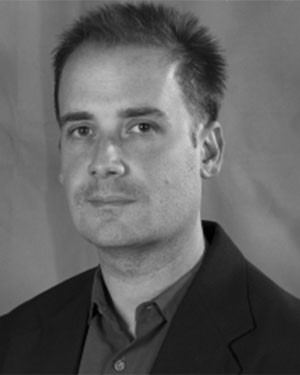}}]{Jonathan Woodring}
received the PhD degree in computer science from Ohio State University in
2009, his specialization in computer graphics and scientific visualization. 
He is a research scientist at the Los Alamos National Laboratory. 
His research areas focus on data science at scale, scientific supercomputing, the intersection of high-performance and cloud computing, future power grids, and ocean climate modeling.
\end{IEEEbiography}

\begin{IEEEbiography}[{\includegraphics[width=1in,height=1.25in,clip,keepaspectratio]{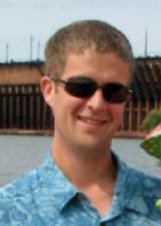}}]{Luke P. Van Roekel}
is currently a research scientist in the climate, ocean, and sea-ice modeling group at LANL and is the co-lead developer of the MPAS-Ocean model, the ocean component of the DOE's new Exascale EnergyEarth System Model (E3SM) and is a science focus group co-lead on the E3SM project. Some of his research interests include ocean surface boundary layer turbulence, large scale earth system dynamics, ocean-atmosphere coupling phenomena, and parameterization of these processes for earth system models. Before joining LANL he was an assistant professor of atmospheric science at a liberal arts college in Wisconsin.
\end{IEEEbiography}

\begin{IEEEbiography}[{\includegraphics[width=1in,height=1.25in,clip,keepaspectratio]{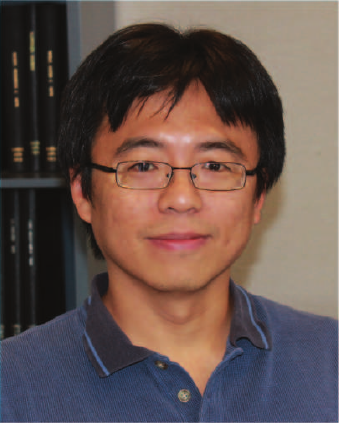}}]{Han-Wei Shen}
is a full professor at the Ohio State University. He received his B.S. degree from the Department of Computer Science and Information Engineering at National Taiwan University in 1988, his M.S. degree in computer science from the State University of New York at Stony Brook in 1992, and his Ph.D. degree in computer science from the University of Utah in 1998. From 1996 to 1999, he was a research scientist at NASA Ames Research Center in Mountain View California. His primary research interests are scientific visualization and computer graphics.  He is a winner of the National Science Foundation's CAREER award and U.S. Department of Energy's Early Career Principal Investigator Award. He also won the Outstanding Teaching award twice in the Department of Computer Science and Engineering at the Ohio State University.
\end{IEEEbiography}

\end{document}